\documentclass[sigconf]{acmart}

\usepackage{multirow}
\usepackage{enumitem}
\usepackage{xspace}
\usepackage{makecell}
\usepackage{acmart-taps}
\usepackage{listings}
\usepackage{framed}

\newcommand{\rev}[1]{#1}

\newcommand{\tool}[0]{\textsf{AMBIPOM}\xspace}

\newcommand{\eg}{{\it e.g.}}
\newcommand{\ie}{{\it i.e.}}

\definecolor{promptTitleBg}{HTML}{2C3E50}
\definecolor{promptBodyBg}{HTML}{F8F9FA}

\lstdefinestyle{promptcontent}{%
    basicstyle=\scriptsize\ttfamily,
    breaklines=true,
    breakatwhitespace=false,
    breakindent=0pt,
    columns=fullflexible,
    keepspaces=true,
    upquote=true,
    showstringspaces=false,
    aboveskip=0pt,
    belowskip=0pt,
    xleftmargin=8pt,
    xrightmargin=8pt,
    linewidth=\dimexpr\linewidth-16pt\relax,
    resetmargins=true,
    literate=
      {•}{{-}}1
      {–}{{-}}1
      {—}{{-}}1
      {“}{{"}}1
      {”}{{"}}1
      {‘}{{'}}1
      {’}{{'}}1
      {…}{{...}}3
      {→}{{->}}2
      {≤}{{$\le$}}1
      {≥}{{$\ge$}}1
      {×}{{$\times$}}1
}

\makeatletter
\newenvironment{promptbox}[2][]{%
    \par\smallskip%
    \begingroup
    \setlength{\FrameRule}{0.5pt}%
    \setlength{\FrameSep}{0pt}%
    \setlength{\OuterFrameSep}{0pt}%
    \MakeFramed{\advance\hsize-\width\FrameRestore}%
    \noindent
    {\setlength{\fboxsep}{0pt}%
    \colorbox{promptTitleBg}{%
        \makebox[\linewidth][l]{%
            \hspace{6pt}%
            \rule[-5pt]{0pt}{3.4ex}%
            \normalfont\bfseries\footnotesize\color{white}#2%
        }%
    }}%
    \if\relax\detokenize{#1}\relax\else\label{#1}\fi%
    \par\vspace{4pt}
    \ignorespaces
}{%
    \par\vspace{5pt}
    \endMakeFramed
    \endgroup
    \par\vspace{2pt}
}
\makeatother

\newcommand{\promptdivider}{%
  \par\vspace{0pt}
  \noindent
  \leaders\hbox{\rule{4pt}{0.35pt}\hskip 3pt}\hfill\kern0pt
  \par\vspace{6pt}
}

\AtBeginDocument{%
  }

\acmYear{2026}\copyrightyear{2026}
\setcopyright{cc}
\setcctype{by}
\acmConference[ACM CAIS '26]{ACM Conference on AI and Agentic Systems}{May 26--29, 2026}{San Jose, CA, USA}
\acmBooktitle{ACM Conference on AI and Agentic Systems (ACM CAIS '26), May 26--29, 2026, San Jose, CA, USA}
\acmDOI{10.1145/3786335.3813144}
\acmISBN{979-8-4007-2415-2/26/05}

\begin{document}

\title{How to Steer Your Multi-Agent System: Human-LLM Collaborative Planning}

\author{Zeyu He}
\authornote{Work done during internship at Megagon Labs.}
\orcid{0009-0007-7115-2692}
\email{zmh5268@psu.edu}
\affiliation{%
  \institution{Penn State University}
  \city{State College}
  \state{Pennsylvania}
  \country{United States}}

\author{Hannah Kim}
\orcid{0000-0002-0137-7171}
\email{hannah@megagon.ai}
\affiliation{%
  \institution{Megagon Labs}
  \city{Mountain View}
  \state{California}
  \country{United States}}

\author{Dan Zhang}
\orcid{0000-0002-6330-0217}
\email{dan_z@megagon.ai}
\affiliation{%
  \institution{Megagon Labs}
  \city{Mountain View}
  \state{California}
  \country{United States}}

\author{Estevam Hruschka}
\orcid{0000-0003-1499-2808}
\email{estevam@megagon.ai}
\affiliation{%
  \institution{Megagon Labs}
  \city{Mountain View}
  \state{California}
  \country{United States}}

\renewcommand{\shortauthors}{He et al.}

\begin{abstract}
  \rev{In orchestrated multi-agent systems, humans often struggle to manage plans due to their complexity and limited transparency. Existing approaches rely on outcome-level supervision, where users verify only final outputs without visibility into intermediate reasoning. We formalize a design space for human-LLM co-planning interactions along three axes: \emph{mode} (semantic vs. structural), \emph{scope} (global vs. targeted), and \emph{level} (low- vs. high-level edits). We realize it in \tool, a prototype supporting process-level supervision through both semantic and structural interactions. Through a user study, we characterize how users navigate this space, revealing hybrid workflows and effort-control-risk trade-offs; through a controlled benchmark, we analyze how LLMs revise plans under varying scope and revision strategies. Our findings yield design insights for more transparent, controllable, and effective human-AI co-planning. We release code and data at \url{https://github.com/megagonlabs/ambipom}.}
\end{abstract}



\maketitle

\section{Introduction}
In recent years, the paradigm of agentic systems has shifted from single, monolithic agents to multi-agent systems (MAS)~\cite{li2024survey,wang2024survey}. Unlike monolithic architectures, MAS consist of multiple specialized agents, each designed with distinct capabilities or domain expertise, for example, mathematical reasoning agents, information extraction agents, or real estate advisory agents~\cite{wu2024autogen,hong2023metagpt}. This modularity enables MAS to tackle complex tasks more efficiently, exploit parallelism for faster execution, and maintain robustness through fault tolerance when individual agents fail~\cite{tran2025massurvey}.
Within the spectrum of MAS architectures, orchestrated MAS, in which a central controller coordinates the agents, offer additional advantages. The controller plays a crucial role in task decomposition, agent assignment, and orchestration of execution, effectively performing high-level planning across the system. This centralized orchestration facilitates global optimization and coherence, clear accountability and control, and more effective allocation of resources and agent capabilities than fully decentralized MAS~\cite{moore2025taxonomyhierarchicalmultiagentsystems,tran2025massurvey,10610676}. 

Recently, large language models (LLMs) have increasingly been employed as planners within orchestrated MAS thanks to their ability to break down high level tasks into manageable subtasks specialized agents~\cite{huang2024understandingplanningllmagents}. However, relying solely on LLM planners remains insufficient in practice. Generated plans may misalign with human intent, fail to capture domain-specific constraints, or suffer from hallucinations~\cite{valmeekam2023on,huang2024understandingplanningllmagents}. 
Therefore, systematic mechanisms are needed to incorporate user interventions effectively and ensure that generated plans remain both correct and aligned with user goals~\cite{zou2025survey}.

Unfortunately, most existing systems rely on \textit{outcome-based supervision}: users are presented with a final answer or, at best, a linear high-level plan while intermediate states remain opaque. This lack of transparency makes it difficult to diagnose failures and apply targeted corrections, especially for complex multi-agent plans.
These limitations call for \textit{process-oriented supervision}~\cite{reppert2023iterateddecompositionimprovingscience} in MAS, where humans can observe dependencies among agent subtasks, evaluate individual outputs, and intervene at multiple stages of planning and execution. By making intermediate reasoning and coordination explicit, process supervision fosters trust and improves controllability in complex MAS.

To operationalize process-level supervision in orchestrated MAS, we adopt a \textit{human-LLM interactive co-planning paradigm}~\cite{horvitz1999principles}.
We represent multi-agent plans explicitly as directed acyclic graphs (DAGs), where nodes correspond to agent-executable subtasks and edges encode data dependencies~\cite{kim2025aipom}. This representation serves as a first-class planning object that can be inspected and modified prior to and during execution.
\rev{Building on this representation, we formalize a \textit{co-planning interaction design space} along three axes:
\textbf{mode}---structural direct manipulation of the graph vs.\ semantic natural-language feedback;
\textbf{scope}---global feedback affecting the entire plan vs.\ targeted feedback conditioned on a selected subgraph; and
\textbf{level}---low-level graph edits (add/edit/delete) vs.\ high-level structural operations (merge/split/replan).
By making \emph{scope} and \emph{level} first-class alongside \emph{mode}, our framework introduces interaction types beyond those exposed by prior co-planning systems and makes different human-LLM collaboration strategies formally comparable. 
We realize this design space in \tool (\textbf{A}gent-aware \textbf{M}ixed-initiative \textbf{B}lock-level \textbf{I}nteractive \textbf{P}lanning for \textbf{O}rchestrated \textbf{M}ulti-agent systems), a prototype to study user and LLM behavior across this space.}

Our contributions are threefold.
\rev{First, we formalize a design space for human-LLM co-planning along three axes (\emph{mode}, \emph{scope}, \emph{level}) and instantiate it in \tool, making different co-planning strategies explicit, operationalizable, and comparable.  
Second, through a user study, we examine how users combine these interaction types to steer and refine plans.
Third, through a controlled benchmark, we analyze the capabilities and limitations of LLMs as co-planning partners across these axes.}
\label{sec:intro}

\section{Related Work}
\subsection{Multi-Agent Systems}
Recent progress in large language models (LLMs) has shifted the focus from single-agent architectures toward multi-agent systems (MAS) for solving complex and long-horizon tasks~\cite{sun2025multiagentcoordinationdiverseapplications}. Instead of relying on a single agent to perform reasoning, planning, and tool use within one loop, MAS distribute these responsibilities across specialized agents, improving scalability while reducing the burden on individual agents~\cite{tran2025massurvey,ijcai2024p890,talebirad2023multiagentcollaborationharnessingpower}.

Among coordination strategies, centralized or hierarchical orchestration has emerged as a practical design pattern~\cite{tran2025massurvey,moore2025taxonomyhierarchicalmultiagentsystems}. In this setting, agents are non-proactive and invoked only when assigned specific subtasks by a central controller. Such systems typically adopt a plan-then-execute paradigm (sometimes referred to as decomposition-first or plan-ahead)~\cite{liu-etal-2025-select-decompose,huang2024understandingplanningllmagents}: a central planner interprets high-level user intent and decomposes it into a structured plan. This plan is often represented as a directed acyclic graph, where nodes correspond to subtasks and edges encode execution order or data dependencies. Execution follows this predefined plan.
Compared to decentralized approaches, centralized orchestration provides explicit data flow, reduced communication overhead, and globally coherent task planning, making it particularly suitable for human-in-the-loop settings, where transparency and controllability are essential.

\subsection{Human-LLM Co-Planning}
Recent advances in LLMs have demonstrated remarkable capabilities in high-level planning and reasoning, making them as promising planners for coordinating MAS without extensive task-specific training~\cite{wei2022chain,huang2024understandingplanningllmagents}. Despite these strengths, fully autonomous LLM planners encounter practical limitations, especially in domain-specific or high-stakes scenarios, due to issues such as hallucination and misalignment with human preferences and expertise~\cite{kambhampati2024position,valmeekam2023on,huang2025survey}. These challenges motivate a mixed-initiative human-in-the-loop (HITL) approach, where human guidance is leveraged to refine, validate, or modify generated plans~\cite{zou2025survey}.

In mainstream AI assistants, including GitHub Copilot and Gemini Deep Research, planning typically occurs within a conversational interface. Within this paradigm, humans function as passive consultants~\cite{feng2025levelsautonomyaiagents}, \ie, reviewing generated plans and requesting changes via high-level prompts. However, this chat-based approach lacks the transparency and granular control required to manage complex multi-agent plans. Addressing this, recent frameworks~\cite{mozannar2025magentic,feng2026cocoa,kim2025aipom,shao2026collaborative} shift the human's role to an active collaborator~\cite{feng2025levelsautonomyaiagents,10.1145/3706598.3713581} through interactive co-planning, co-execution, and debugging.
Frameworks like COCOA~\cite{feng2026cocoa} and Magnetic-UI~\cite{mozannar2025magentic} allow humans to directly manipulate linear plans by adding, editing, or removing plan steps. Alternatively, AGDebugger~\cite{10.1145/3706598.3713581} enables humans to send or edit messages sent among agents including the orchestrator. 
\rev{The closest prior work, AIPOM~\cite{kim2025aipom}, combines chat with a graph editor and shares our goal of process-level supervision, but realizes it along the \emph{mode} axis only. Our work is the first to formalize a co-planning interaction design space along three axes (\emph{mode}, \emph{scope}, \emph{level}), populating them with new interaction types and enabling controlled comparison.}
\label{sec:related}

\section{Problem Statement}
Given a task query $Q$ and a set of agent descriptions $A$, a planner generates a plan $P$ for a multi-agent system (MAS).
\rev{We formalize human-LLM collaborative planning along three axes: (1) \textbf{mode}: structural interactions, where the user directly manipulates the plan graph, vs.\ semantic interactions, where the user provides high-level feedback interpreted by the planner; (2) \textbf{scope}: global interactions affecting the entire plan vs.\ targeted interactions modifying a selected subgraph; and (3) \textbf{level}: low-level edits (atomic graph operations) vs.\ high-level edits (compositional operations such as merge or split).}
Formally, the plan update can be represented as
\[
P' = \text{update\_plan}\big(P_0, (\text{mode}, \text{scope}, \Delta, S_0 \text{ if scope = targeted}))\big),
\]
where 
\(\text{mode} \in \{\text{structural}, \text{semantic}\}\),  
\(\text{scope} \in \{\text{global}, \text{targeted}\}\),  
\(\Delta\) corresponds to either direct manipulation on a graph or text feedback, and  
\(S_0 \subseteq P_0\) is the selected subgraph for targeted interactions.

\[
\text{update\_plan} =
\begin{cases} 
\text{apply\_ops}(P_0, \Delta_{DM}) & \text{if structural} \\[2mm]
\text{planner}(P_0, \Delta_{Text}) & \text{if semantic / global} \\[1mm]
\text{planner}(S_0, \Delta_{Text}) \;\oplus\; (P_0 \setminus S_0) & \text{if semantic / targeted}
\end{cases}
\]

\subsection{Research Questions}\label{sec-rq}
Our study investigates two complementary aspects of collaborative planning in MAS: how users interact with the system to steer and refine plans (RQ1,2) and how LLM-based planners revise structured plans under varying feedback scopes and strategies (RQ3,4).

\begin{itemize}[leftmargin=1.5em]
    \item \textbf{RQ1 (User Performance with Interaction Types)}
	How do interaction types--especially targeted semantic interactions and high-level structural interactions--affect user efficiency, effort, and task success compared to baseline interactions?
    \item \textbf{RQ2 (User Preference and Strategy)} 
	How do users select among interaction types depending on their plan modification goals, and what patterns emerge in their human–LLM collaboration strategies?
    \item \textbf{RQ3 (Effect of Feedback Scope)}
    How does the scope of feedback (targeted vs. global) affect the quality of LLM-generated plan revisions in terms of edit correctness and global coherence?
    \item \textbf{RQ4 (LLM Revision Strategies)}
	How do different LLM revision strategies--direct plan regeneration vs. edit-sequence (structural operation) generation--compare in terms of correctness, interpretability, and robustness?
\end{itemize}

\subsection{Design Goals}\label{section-dg}
Motivated by our research questions, we define three design goals for human–LLM collaborative planning systems that emphasize transparency and controllability. 
\rev{Such systems should support flexible interactions across mode, scope, and level} 
while preserving plan integrity~\cite{horvitz1999principles,ai2004mapgen,chen2025dango,freund2025flowco}. Building on these principles, we highlight three key goals that address gaps in existing systems.

\textbf{DG1 (Support Targeted and Safe Interactions)}
Planning systems should provide fine-grained control over plan revisions, allowing users to refine specific parts of a plan without affecting the rest~\cite{masson2024directgpt,kim2025aipom}. Preserving boundary interfaces and limiting cascading changes ensures targeted adjustments are safe and predictable, supporting more precise human guidance.

\textbf{DG2 (Support Complex Structural Edits)}
Beyond low-level graph edit operations, such as adding, editing, or removing subtasks, planning systems should facilitate high-level structural modifications such as merging, splitting, or branching that consist of a sequence of low-level edits~\cite{kim2025aipom,freund2025flowco,liu2025interactive,li2023interactive}. Automating or assisting these complex operations reduces user effort and errors, addressing the high friction of structural plan refinement in existing work.

\textbf{DG3 (Preserve Transparency, Traceability and Inspectability)}
Users should be able to inspect node-level execution status, intermediate plan states, and the history of interactions~\cite{wang2025survey,barrak2025traceability}. These capabilities enable iterative refinement, support error diagnosis, and facilitate informed decision-making.

Based on these design goals, we implemented \tool, a prototype human–LLM co-planning system to study human–LLM collaborative behaviors and investigate the research questions outlined above.
\label{sec:problem}

\section{\tool: Human-LLM Collaborative Planning}
\tool consists of an LLM-based planner and four execution agents (LLMs with tools enabled) specialized in code, math, search, and commonsense tasks. Implementation details are provided in Appendix~\ref{appendix:imple}, including prompts for the planner (\S~\ref{appendix:planner_prompts}) and execution agents (\S~\ref{appendix:executor_prompts}).

To facilitate transparency and controllability, \tool comes with a dual-panel interface, featuring a chat panel for semantic interaction and an interactive plan DAG visualization for structural interaction, similar to \citet{kim2025aipom}.

\subsection{LLM-Based Planner}
The planner is an LLM-based module responsible for generating the initial plan and revising the plan in response to user interactions. It supports four interaction types ($\S~\ref{sec:global_feedback}-\ref{sec:llm_assisted_dm}$) spanning different interaction modes (semantic vs. structural), scopes (global vs. targeted), and edit levels (low-level vs. high-level).

\subsubsection{Initial Plan Generation}
The initial plan $P_0$ is generated entirely by the LLM, \ie, $P_0 = \text{planner}(Q, A)$.

\subsubsection{Semantic Interaction: Global Feedback \textbf{(GF)}}
\label{sec:global_feedback}
Users can provide free-form textual feedback $\Delta_{\text{text}}$ that applies to the entire plan. The planner regenerates the plan as:
\begin{equation}
P' = \text{planner}(P_0, \Delta_{\text{Text}})
\end{equation}
This interaction type allows holistic revisions that reflect changes in overall strategy or task goals.

\subsubsection{Semantic Interaction: Targeted Feedback \textbf{(TF)} (Supporting DG1)}
\label{sec:targeted_feedback}
Users can select a subgraph $S_0 \subseteq P_0$ and provide textual feedback $\Delta_{\text{Text}}$ specific to that region. 
The planner then regenerates only the selected subgraph while preserving compatibility with the rest of the plan:
\begin{equation}
P' = \text{planner}(S_0, \Delta_{\text{text}}) \oplus (P_0 \setminus S_0)
\end{equation}
where $\oplus$ denotes reintegration of the revised subgraph into the unchanged portion of the plan. 
To encourage the LLM to follow the prompt template and preserve required interfaces, we intentionally minimize the context provided during replanning.
In particular, the planner is prompted only with the selected subgraph $S_0$ and its boundary input/output specification, rather than the full conversation history and the full plan $P_0$, which can introduce irrelevant details and increase format violations.

\subsubsection{Structural Interaction: Direct Manipulation \textbf{(DM$_{low}$)}}
\label{sec:low_dm}
Users can perform low-level operations (single graph edit operations) on plan DAG, such as adding or deleting nodes or edges, modifying task descriptions, reassigning agents, updating input/output fields, or updating agent configurations, \ie, direct manipulation  (DM). 
These operations are applied deterministically and do not involve the LLM:
\begin{equation}
P' = \text{apply\_ops}(P_0, \Delta_{DM}),
\end{equation}
where $\Delta_{DM}$ represents the set of user-specified direct manipulations on the plan graph.

\begin{figure*}[!tb]
    \centering    \includegraphics[width=\textwidth]{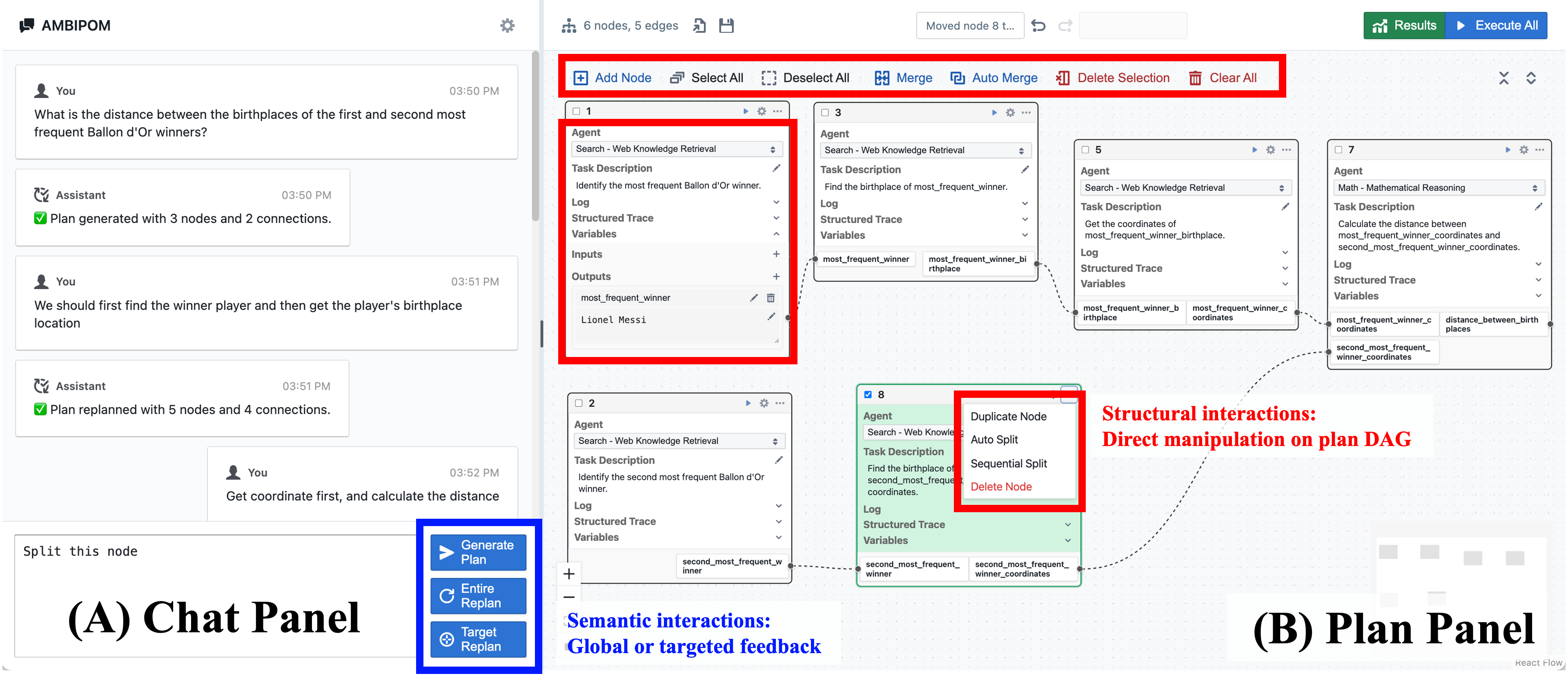}
    \caption{\tool supports transparent and controllable human–LLM co-planning through a dual-panel interface. 
    \textbf{(A) Chat Panel} \rev{supports plan generation, replanning, and execution feedback, with textualized logs of plan changes for transparency.}
    \textbf{(B) Plan Panel} visualizes the current plan as an editable graph, allowing users to inspect and refine the workflow via DMs.
    }
    \label{fig:ui}
\end{figure*}

\subsubsection{Structural Interaction: High-Level DM and LLM-Assistance \textbf{(DM$_{high}$)} (Supporting DG2)}
\label{sec:llm_assisted_dm}
Beyond low-level DM, \tool supports high-level DM (\ie, compositional edit operations that combine multiple low-level operations into a single meaningful interaction) such as merge or split, which can be performed both manually and with LLM assistance.
For manual edits, the merge operation collapses a selected interface-closed subgraph $S_0 \subseteq P_0$ into a single node $v_\text{new}$, preserving the subgraph's external input/output interfaces. 
The split operation divides a single node $v \in P_0$ into two sequential nodes $v_1, v_2$, while maintaining the original input/output connectivity.

In LLM-assisted editing, denoted by \textbf{DM$_{high}^+$}, the system automatically generates a structural edit operation $\Delta_\text{LLM}$ for either auto-merge or auto-split, based on the details of the selected subgraph $S_0$ (auto-merge) or node $v$ (auto-split). The operation is then applied deterministically to $P_0$:
\begin{equation}
\Delta_\text{LLM} = \text{planner}_{op}(\text{selection}), \quad P' = \text{apply\_ops}(P_0, \Delta_\text{LLM}),
\end{equation}
where \text{selection} denotes either a subgraph for an auto-merge or a node for an auto-split. By automating these high-level structural operations, the LLM-assisted edits reduce user effort while maintaining structural validity.

\subsection{Interface (Supporting DG3)}
\tool provides a dual-panel interface centered on a shared plan state, which tracks the plan DAG, node execution status, and user interaction history. 
Updates from either panel are propagated immediately, ensuring consistent views across semantic and structural interactions.
The chat panel serves as the primary channel for natural language interaction and system feedbacks.
The plan panel serves as the primary workspace for inspecting and editing the visualized plan graph through structural interactions.

\subsubsection{Chat Panel (Fig.~\ref{fig:ui}A)}
The chat panel provides a conversation interface semantic interactions, allowing users to issue high-level instructions and receive system status messsages. It supports a range of high-level instructions, including (1) initializing a plan via ``Generate Plan'' and (2) refining an existing plan through ``Entire Replan'' (global replanning) or ``Targeted Replan,'' (subgraph-specific replanning), enabling flexible control over revision scope.

User messages and system responses are displayed as chat bubbles, forming a traceable interaction history. 
During plan generation and replanning, the system posts concise summaries of changes (\ie, the number of nodes/edges created, which nodes were added or removed) to help users understand.
In addition, the chat panel surfaces system activity notifications that record DM activity from the plan panel. For basic DMs, these logs capture the performed actions for activity tracking; for LLM-Assisted DMs, such as merge and split, they also summarize the resulting node-level changes.

\begin{figure}
    \centering
    \includegraphics[width=1\linewidth]{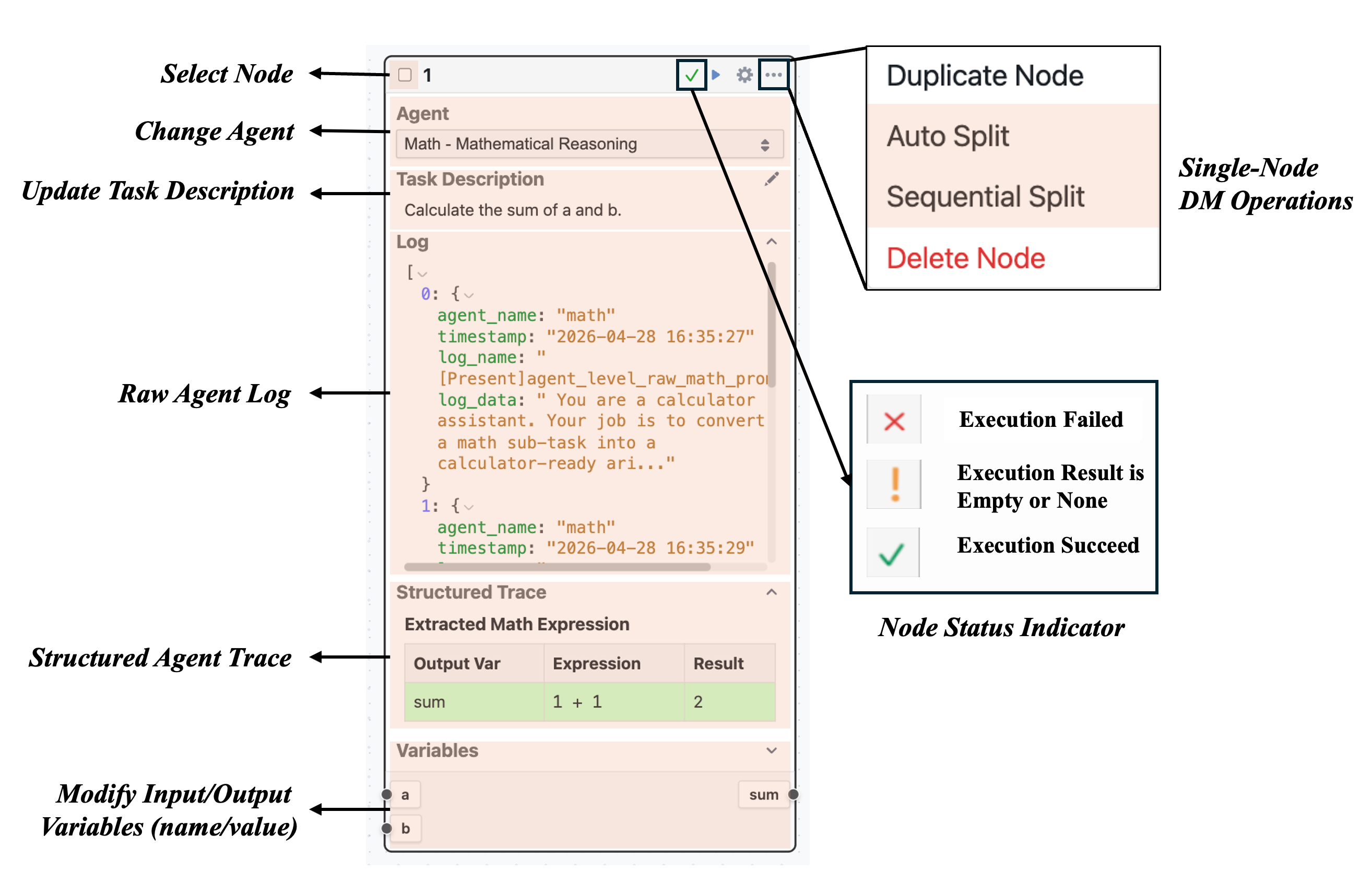}
    \caption{\rev{Each node card shows the agent, task, status, editable I/O fields, and post-execution logs. Selection enables merge; single-node operations via the three-dot menu.}
    }
    \label{fig:node-ui}
\end{figure}

\subsubsection{Plan Panel (Fig.~\ref{fig:ui}B)}
The plan panel visualizes the current plan as an editable DAG, where nodes represent subtasks and edges denote data dependencies. 
Each node (Fig.~\ref{fig:node-ui}) is rendered as a card displaying the subtask description, assigned agent, input/output fields, and node status indicator.
Nodes also expose execution logs in two formats: raw log and structured trace that vary by agent inner pipelines.
For example, the math agent's trace shows generated expressions and computed results; the code agent shows generated code and execution outputs; the search agent shows search queries, retrieved results, and summaries; and the commonsense agent surfaces intermediate reasoning alongside final outputs.

Nodes can be executed individually, with outputs and status updated immediately.
Users can perform direct manipulations (DM), including adding, deleting, duplicating, or repositioning nodes, linking or removing edges, and editing subtask attributes (\eg, agent assignment, input/output variables, task descriptions).
The plan panel also supports high-level structural interactions: interface-closed subgraphs can be merged into a single node, either manually or via LLM-assisted auto-merge, while nodes can be split sequentially or via LLM-assisted auto-split based on task description and I/O fields. Selection shortcuts, undo/redo, and collapse/expand buttons aid iterative plan refinement, while control buttons at top-right allow execution of the full plan (“Execute All”) and viewing aggregated outputs (“Results”).

This design integrates low-level DM and high-level LLM-assisted edits, enabling transparent, efficient, and controllable structural interactions that complement the chat panel’s semantic interactions.
\label{sec:method}

\section{User Study: Understanding Human–LLM Collaboration}
To study how humans interact with LLMs in co-planning, we conducted a controlled user study comparing two system prototypes. We used \tool, our prototype human–LLM co-planning system, and a baseline version implemented with standard interactions offered in prior works (\textbf{GF}, \S~\ref{sec:global_feedback}, \textbf{DM$_{low}$}, \S~\ref{sec:low_dm}). To mitigate confounding effects of interface design, the baseline retained a user interface similar to \tool.

While \tool includes several features to support co-planning, the primary difference relevant to our research questions was the availability of advanced interaction types: \textbf{targeted semantic feedback} (\textbf{TF}, \S~\ref{sec:targeted_feedback}) and \textbf{high-level structural edit} (\textbf{DM$_{high}$}, \S~\ref{sec:llm_assisted_dm}). We employed a within-subject design in which participants completed two sessions (one per system), allowing direct comparison of task efficiency, effort, and success across interaction types. 
This design allowed us to examine the effects of interaction types on performance (RQ1) and how users select and combine them to steer plans (RQ2).

\subsection{Study Design}

\subsubsection{Participants}
We recruited 13 voluntary participants (P1–P13) in-house from a US-based industry research lab using convenience sampling.
A background survey revealed that 69\% of participants reported using LLMs constantly, while 31\% reported daily use. 
On 7-point Likert scales, participants’ familiarity with LLMs was high (M = 5.92, SD = 0.95), as was their awareness of LLM limitations (M = 5.62, SD = 0.96). Trust in LLM-generated responses was below the scale midpoint (M = 3.62, SD = 1.04), whereas verification habits were higher (M = 5.31, SD = 1.18), indicating that participants tended to critically assess LLM outputs despite frequent use.

\subsubsection{Tasks}\label{sec:user-study-task}
We curated four sets of question pairs (Table~\ref{tab:user-study-question}), with each pair designed to have similar difficulty and planning complexity. The questions were crafted to involve multiple agents and could not be solved by a single agent alone, requiring a combination of numerical computation, coding, retrieval, and commonsense reasoning. Each pair belongs to one of four distinct structural patterns:
\begin{enumerate}[leftmargin=1.5em,topsep=0pt]
    \item Stepwise Math Reasoning: multi-step arithmetic with no retrieval (\eg, value increase when flipping an item).
    \item Multi-Hop Computation: chained retrievals leading to comparative computation (\eg, distance between MVPs’ birthplaces).
    \item Listed Retrieval \& Aggregation: list given in the query; retrieve per item and aggregate (\eg, total chapters in a book series).
    \item Top-K Retrieval \& Aggregation: discover a top-K list, retrieve per item, and aggregate (\eg, average top-university tuition).
\end{enumerate}
For each question, an initial plan was provided, which could contain different kinds of errors. Participants’ task was to revise the plan using the system, and produce the correct final answer.
Each participant completed all eight questions in two sessions of four questions each, with one question from each structural pattern per session. Question order and error type assignment were randomized to mitigate learning and order effects.

\subsubsection{Procedure}
Each study was conducted in person and lasted between 70 to 171 minutes. 
With consent, screen, audio, and video were recorded for analysis.

Participants first completed a short background survey on LLM usage and familiarity.
The main part consisted of two sessions, one with each system (\tool and the baseline). To mitigate order effects, system order was counterbalanced: seven participants used \tool first, and six participants used the baseline first.
In each session, participants first watched a brief demonstration and completed a tutorial task to familiarize themselves with the interface, followed by four study tasks (one per question type).
After finishing the four tasks, participants completed post-session questionnaires, including SUS and NASA-TLX.
Participants then proceeded to a second session with the other system, following the same structure (demo, tutorial task, four tasks, and questionnaires).
Finally, participants completed a post-study survey that captured system preferences and open-ended feedback.

\subsubsection{Metrics}\label{perform-metrics}
We measure task performance along two dimensions: effectiveness (task success, plan completeness) and efficiency (completion time, interaction frequency, conversation turns). Detailed definitions in Appendix~\ref{appendix:user-study-metrics}.

\subsection{User Study Results}
In this section, we summarize our findings under user-side research questions (RQ1,2) mentioned in \S~\ref{sec-rq}. Detailed analyses are included in Appendix~\ref{appendix:user-study-detail-results}.\footnote{Parts of this summary, derived from the detailed analyses in Appendix~\ref{appendix:user-study-detail-results}, were initially generated using ChatGPT and Gemini and subsequently edited by the authors for clarity and accuracy.}

\subsubsection{RQ 1: Performance with Advanced Interaction Types}\label{sec:rq-1-discussion}
We compared how advanced interactions (\textbf{TF, DM$_{high}$}) affected plan quality and efficiency compared to basic interactions (\textbf{GF, DM$_{low}$}). While we hypothesized that advanced types would produce better plans (H1), participants actually produced slightly higher-quality plans using the basic types. However, advanced features did provide distinct usability benefits: \textbf{DM$_{high}$} reduced users' cognitive load compared to manual \textbf{DM$_{low}$}, and targeted text feedback \textbf{TF} successfully reduced the number of conversational turns required to refine a plan (H3). Across both systems, semantic interactions generally reduced mental effort, while structural interactions were perceived to match users' exact intentions more closely.

\rev{To disentangle the contributions of \textbf{TF} and \textbf{DM\textsubscript{high}\textsuperscript{+}} in participants' refinement workflows, we grouped each participant-question observation by which advanced types appeared: \textbf{TF}+\textbf{DM\textsubscript{high}\textsuperscript{+}}, \textbf{TF} only, \textbf{DM\textsubscript{high}\textsuperscript{+}} only, and (for observations using neither) \textbf{GF}-dominant or \textbf{DM\textsubscript{low}}-dominant workflows. Following the effectiveness metrics in Appendix~\ref{appendix:user-study-metrics}, we compared these groups on final-answer accuracy and plan quality.
The results suggest that \textbf{DM$_{high}^+$} was associated with better plans and higher accuracy than \textbf{TF}. This aligns with post-study preference for \textbf{DM\textsubscript{high}\textsuperscript{+}} and with \textbf{TF}'s higher application difficulty: users need to select the correct subplan boundary and formulate targeted textual feedback.
It is also consistent with our benchmark experiment (\S\ref{sec:exp:results}), where \textbf{TF} underperforms more constrained revision strategies. 
This refines our earlier finding: within advanced interaction types, \textbf{DM\textsubscript{high}\textsuperscript{+}} is the more beneficial component, while \textbf{TF} introduces greater difficulty and variability.}

\subsubsection{RQ 2: User Strategy and Collaboration Patterns}
To understand type selection during collaborative refinement, we triangulated 
(1) post-task preference ratings,
(2) interaction logs,
and (3) think-aloud and observational notes across 13 participants.
Participants could refine plans via four types: manual \textbf{DM}, LLM-assisted \textbf{DM$_{high}^+$}, \textbf{GF} (full-plan replanning), and \textbf{TF} (subplan replanning).
In addition, we coded participants' text feedback into two functional categories: prescriptive/instructive (specifying corrective steps or transformations) and descriptive/diagnostic (identifying an error or mismatch and requesting correction). 

\paragraph{Users did not ``pick a type''; they assembled hybrid workflows per iteration:} Rather than sticking to a single interaction type, users dynamically alternated between modes and scopes. A typical sequence involved using text feedback for broad modifications and then switching to manual \textbf{DM} for fine-grained, local repairs.
\paragraph{Intentions for structural modification pulled strongly toward LLM-assisted \textbf{DM$_{high}^+$}; semantic modification were more mixed:} When making graph-level changes, users heavily favored LLM-assisted \textbf{DM$_{high}^+$} (\eg, auto-merge or auto-split) for rapid restructuring. For semantic tweaks, preferences were split, and users frequently paired targeted text feedback (\textbf{TF}) with manual \textbf{DM} to safely finalize adjustments.
\paragraph{Type choice was mediated by an effort-control-risk trade-off:} Interaction choices were driven by balancing manual effort against the risk of the LLM breaking the plan. Users treated manual \textbf{DM} as high-control but high-effort, global text feedback (\textbf{GF}) as low-effort but risky (like ``rolling a dice''), and LLM-assisted \textbf{DM$_{high}^+$} as a favored middle ground.
\paragraph{Collaborative refinement showed two recurring rhythms:} review-first versus execute-first: Most participants (10/13) preferred to thoroughly inspect and clean the plan before executing it, while a few (3/13) executed early to let failures guide their edits. Regardless of their rhythm, both groups converged on a similar iterative cycle: broad text edits, local DM patches, and then re-execution.
\paragraph{Users used text feedback as an executable specification, switching to diagnostic feedback when debugging:} Text feedback was overwhelmingly prescriptive (169 out of 178 messages). Users primarily commanded the LLM like an algorithmic executor (\eg, ``split this node''), and only resorted to diagnostic feedback (describing errors like a bug report) when actively debugging a breakdown.
\paragraph{Trust increased in-task but verification declined later in sessions (The Trust-Fatigue Paradox):} While users' trust in LLM-assisted features (\textbf{DM$_{high}^+$}) grew as they gained experience with the system, their rigorous verification habits decayed over time. Due to fatigue, users shifted from carefully checking intermediate results early in the session to accepting plausible-looking outputs without verification later on.
\label{sec:user-study}

\section{Experiments: LLM Plan Revision with Feedback}
We evaluate the effectiveness of LLM planners in interpreting and incorporating human feedback to revise multi-agent plan graphs. 
\rev{Our main experiments systematically vary three factors: feedback scope (global vs.\ targeted), context availability (whether the full plan is provided), and revision strategy (direct plan regeneration vs.\ edit-sequence generation). Boundary flexibility (whether boundary interfaces are frozen or can be updated) is studied in Appendix~\ref{sec:additional-boundary-analysis}.
We compare four conditions (GF, TF, TF+P, GF-to-DM) on a benchmark with ground-truth plans and report structural/semantic similarity, stability, and execution accuracy where applicable.}

\subsection{Experimental Setup}

Our experimental design follows a structured ablation approach.
We group conditions by the \emph{revision strategy} they use and then isolate mechanisms within each group.

\subsubsection{Direct Plan Regeneration (GF, TF, TF+P)}
Direct Plan Regeneration conditions use textual feedback to regenerate plans, but vary in two mechanisms that shape the trade-off between local correctness and global coherence: 
feedback scope and context availability.
\textbf{GF} corresponds to Global Feedback that replans the plan with full-plan context and does not require explicit boundary handling.
\textbf{TF} corresponds to Targeted Feedback that replans only on the selected subplan with no full plan context and freezes boundary interfaces during reintegration.
\textbf{TF+P} extends \textbf{TF} by providing the full plan as context during subplan revision, while still freezing boundary interfaces during reintegration.
\rev{To evaluate LLM-Assisted \textbf{DM$_{high}^+$}, we instantiate auto-merge and auto-split within the \textbf{TF} setting (\textbf{TF$_{merge}$} and \textbf{TF$_{split}$}).}

\subsubsection{Edit-Sequence Generation (GF-to-DM)}
\textbf{GF-to-DM} takes a full plan and text feedback, but differs from \textbf{GF} in revision strategy: instead of regenerating a plan using text feedback, the LLM generates a sequence of structural edit operations that will be applied to the original plan to transform it to a new plan.

\subsection{Benchmark Dataset}\label{sec:bench-dataset}
\rev{We construct a benchmark of 200 gold plans and 1,150 broken-plan items with ground-truth structures via reverse operation.}
For each item, we start from a gold plan $p_{\text{gold}}$, apply a breaking operation $f$ to obtain an initial plan $p_{\text{initial}} = f(p_{\text{gold}})$, and generate a natural-language feedback describing the inverse transformation.
Each benchmark item is a tuple $(p_{\text{initial}}, \text{feedback}, \text{target\_nodes}, p_{\text{gold}})$, where \texttt{target\_nodes} specifies the target subgraph to be edited for targeted conditions.

The benchmark is designed to mirror the structural demands of our user study tasks (\S~\ref{sec:user-study-task}) rather than their topics, spanning four subsets: Stepwise Math Reasoning, Multi-Hop Computation, Listed Retrieval \& Aggregation, and Top-K Retrieval \& Aggregation.
For Stepwise Math Reasoning, we select 50 math problems from GSM8K~\cite{cobbe2021gsm8k} whose gold plans are known to execute correctly under our planner and satisfy plan-generation constraints, ensuring that execution-based evaluation is meaningful.
For the remaining three subsets, we use Claude Sonnet 4.6 to draft 50 questions per subset that match the target structural pattern, and manually review them, discarding ambiguous, unsolvable, or off-pattern items.
Across all selected items, we enforce a minimum plan-complexity threshold (at least 5 nodes and 5 edges) to ensure that the reverse-breaking operation and its corresponding inverse instruction are feasible. 

We cover seven operation types: add node, change node description, change node agent, merge (sequential), merge (parallel), split (sequential), and split (parallel).
Split operations require specific node patterns (\eg, a plan with a compatible node structure) and thus appear only in compatible subsets.
In our benchmark, sequential split instances occur only in the Multi-Hop Computation and Top-K Retrieval \& Aggregation subsets, while parallel split instances occur only in the Top-K Retrieval \& Aggregation subset.

\subsection{Experimental Procedure}\label{sec:exp-procedure}
For each benchmark item, we provide an initial plan $p_{\text{initial}}$ and a natural-language feedback describing the intended refinement.
To mirror the interaction modes in our system, we render the feedback differently across conditions: \textbf{global} conditions (\textbf{GF} and \textbf{GF-to-DM}) receive \emph{ID-anchored} instructions that reference specific node IDs, whereas \textbf{targeted} conditions receive \emph{deictic} instructions that intentionally avoid node IDs and instead refer to the selected region.
Each condition then differs in the plan context provided to the LLM.

For \textbf{GF}, the model receives the full plan and outputs a revised full plan.
For targeted conditions, the model receives only the selected subplan (with full-plan context provided in \textbf{TF+P}) and outputs a revised subplan, which is reintegrated into the original plan.
Reintegration checks boundary interface compatibility; if the revised subplan cannot be integrated into the original plan due to a boundary interface mismatch, this run is marked as a failure.
For LLM-Assisted \textbf{DM$_{high}^+$}, we do not provide additional instruction beyond operations because the target node(s) are already selected.
The system applies auto-merge/auto-split within the same context setting as \textbf{TF}.
For edit-sequence-based revision (\textbf{GF-to-DM}), the model receives the full plan and an ID-anchored instruction, but outputs edit sequence operations rather than a revised plan. The system applies operations step-by-step, and any invalid edit-sequence structural operation is treated as a failure. 
We evaluate \textbf{GF-to-DM} only on the Stepwise Math Reasoning subset, and we additionally execute the refined plans produced by both \textbf{GF} and \textbf{GF-to-DM} to compute execution accuracy.

\rev{To complement the synthetic benchmark, we evaluate on naturally-occurring faulty plans (Appendix~\ref{app:natural-faulty-plans}), selected from the outputs of weaker planner models on the benchmark queries.}

\begin{table}[t]
\centering
\small
\setlength{\tabcolsep}{4pt}
\caption{Integration success and failure counts by feedback condition. GF bypasses integration by regenerating the full plan; targeted conditions and GF-to-DM may fail at boundary or operation-validity checks.}
\label{tab:integration-results-no-boundary}
\begin{tabular}{l cccc cc}
\toprule
 & \textbf{GF} & \textbf{TF} & \textbf{TF\textsubscript{split}} & \textbf{TF\textsubscript{merge}}  & \textbf{TF+P}  & \textbf{GF-to-DM}  \\
\midrule
\textbf{Success}   & 1150& 1122& 150&397 & 1134  & 197      \\
\textbf{Failure}    & 0& 28& 0&3 & 16 &  53     \\
\midrule
\textbf{Success Rate} & 1& 0.976& 1&0.992 & 0.986 & 0.788           \\
\bottomrule
\end{tabular}
\end{table}

\subsection{Evaluation Metrics}
\rev{We evaluate each refined plan $p_{\text{refined}}$ against $p_{\text{gold}}$ using four quality metrics, computed only over runs that pass an integration check:
(1) \textbf{Graph Edit Distance (GED$\downarrow$)}: structural similarity between plan topologies (0 = exact match).
(2) \textbf{Semantic Similarity (SS$\uparrow$)}: similarity aggregated over node descriptions (range 0-1).
(3) \textbf{Plan Stability (Stable$\uparrow$)}: fraction of non-target nodes unchanged after revision (range 0-1; higher = fewer unexpected side effects).
(4) \textbf{Execution Accuracy $\uparrow$}: applicable only to the Stepwise Math Reasoning subset; we execute revised plans and compute final-answer accuracy for \textbf{GF} and \textbf{GF-to-DM}.}

\rev{The integration check fails for targeted conditions when the revised subplan cannot be integrated due to boundary interface mismatch, and for \textbf{GF-to-DM} when an invalid structural operation is generated; integration success rates are reported in Table~\ref{tab:integration-results-no-boundary}.}

\subsection{Experiment Results}\label{sec:exp:results}

Our findings below address how feedback scope affects revisions (RQ 3: \textbf{GF}, \textbf{TF}, \textbf{TF+P}) and how revision strategies compare (RQ 4: \textbf{GF} vs. \textbf{GF-to-DM}).

\paragraph{Integration failures concentrate in targeted feedback and edit-sequence revisions}
\label{sec:gf-most-failure-resistant}
Table~\ref{tab:integration-results-no-boundary} presents the integration performance across all interaction conditions. The detailed breakdown is shown in Table~\ref{tab:success-rates}.
Global feedback (\textbf{GF}) produced valid revisions in all cases, largely because it is designed to regenerate the entire plan directly, thereby avoiding explicit subgraph reintegration. 
In contrast, targeted conditions exhibited lower success rates because revising only a subgraph can introduce interface mismatches with the unchanged remainder of the plan at reintegration boundaries.
Within the targeted conditions, \textbf{TF+P} has a slightly higher success rate than \textbf{TF}. 
This pattern suggests that full-plan context improves reintegration success.
\rev{Auto-merge/split variants under \textbf{TF} show near-perfect integration.
In contrast, edit-sequence structural operation refinement (\textbf{GF-to-DM}) had the lowest success rate, since later operations depend on accurately tracking the plan state produced by earlier edits.}

\begin{table*}[ht]
\centering
\setlength{\tabcolsep}{4pt}
\renewcommand{\arraystretch}{1.15}
\caption{Plan refinement performance by operation type and feedback condition, measured by graph edit distance (GED$\downarrow$), semantic similarity (SS$\uparrow$), and plan stability (Stable$\uparrow$). Bold marks the best per column. Boundary-flexible variants (+B) are reported in Appendix~\ref{sec:additional-boundary-analysis}.}
\resizebox{\textwidth}{!}{
\begin{tabular}{l|ccc|ccc|ccc|ccc|ccc|ccc|ccc}
\toprule
\multirow{2}{*}{\textbf{Operation}} 
& \multicolumn{3}{c|}{\textbf{Add Node}} 
& \multicolumn{3}{c|}{\textbf{Change Desc.}} 
& \multicolumn{3}{c|}{\textbf{Change Agent}} 
& \multicolumn{3}{c|}{\textbf{Merge Sequential}} 
& \multicolumn{3}{c|}{\textbf{Merge Parallel}} 
& \multicolumn{3}{c|}{\textbf{Split Sequential}} 
& \multicolumn{3}{c}{\textbf{Split Parallel}} \\
\cmidrule(lr){2-4}\cmidrule(lr){5-7}\cmidrule(lr){8-10}\cmidrule(lr){11-13}\cmidrule(lr){14-16}\cmidrule(lr){17-19}\cmidrule(lr){20-22}
& \textbf{GED$\downarrow$} & \textbf{SS$\uparrow$} & \textbf{Stable$\uparrow$} 
& \textbf{GED$\downarrow$} & \textbf{SS$\uparrow$} & \textbf{Stable$\uparrow$} 
& \textbf{GED$\downarrow$} & \textbf{SS$\uparrow$} & \textbf{Stable$\uparrow$} 
& \textbf{GED$\downarrow$} & \textbf{SS$\uparrow$} & \textbf{Stable$\uparrow$} 
& \textbf{GED$\downarrow$} & \textbf{SS$\uparrow$} & \textbf{Stable$\uparrow$} 
& \textbf{GED$\downarrow$} & \textbf{SS$\uparrow$} & \textbf{Stable$\uparrow$} 
& \textbf{GED$\downarrow$} & \textbf{SS$\uparrow$} & \textbf{Stable$\uparrow$} \\
\hline
GF         & \textbf{0.570} & \textbf{0.993} & 0.991 & 0.030 & \textbf{0.994} & 0.981 & 0.005 & \textbf{1.000} & 0.998 & 0.070 & \textbf{0.995} & 0.977 & \textbf{0.375} & \textbf{0.993} & 0.973 & \textbf{0.000} & \textbf{0.999} & 0.984 & \textbf{0.000} & 0.995 & 1.000 \\
TF        & 1.920 & 0.979 & \textbf{1.000} & 0.185 & 0.987 & \textbf{1.000} & 0.620 & 0.993 & \textbf{1.000} & 0.270 & 0.986 & \textbf{1.000} & 1.755 & 0.987 & \textbf{1.000} & 0.020 & 0.995 & \textbf{1.000} & 0.160 & 0.994 & \textbf{1.000} \\
TF$_{merge}$ & - & - & - & - & - & - & - & - & - & \textbf{0.000} & 0.987 & \textbf{1.000} & 1.767 & 0.988 & \textbf{1.000} & - & - & - & - & - & - \\
TF$_{split}$ & - & - & - & - & - & - & - & - & - & - & - & - & - & - & - & 0.370 & 0.993 & \textbf{1.000} & 0.400 & 0.993 & \textbf{1.000} \\
TF+P        & 1.549 & 0.979 & \textbf{1.000} & 0.590 & 0.988 & \textbf{1.000} & 0.935 & 0.982 & \textbf{1.000} & 0.195 & 0.987 & \textbf{1.000} & 1.005 & 0.988 & \textbf{1.000} & \textbf{0.000} & \textbf{0.999} & \textbf{1.000} & 0.080 & 0.994 & \textbf{1.000} \\
\hline
GF-to-DM        & 1.000 & 0.985 & \textbf{1.000} & \textbf{0.020} & 0.989 & \textbf{1.000} & \textbf{0.000} & \textbf{1.000} & \textbf{1.000} & 0.300 & 0.985 & \textbf{1.000} & 1.360 & 0.989 & \textbf{1.000} & - & - & - & - & - & - \\
\hline
\multicolumn{22}{c}{\textbf{Flexible Boundary}}\\
\hline
TF+B        & 2.370 & 0.978 & 0.996 & 0.770 & 0.989 & 0.996 & 0.875 & 0.992 & \textbf{1.000} & 0.930 & 0.987 & \textbf{1.000} & 1.915 & 0.988 & 0.994 & 0.180 & 0.993 & \textbf{1.000} & 17.180 & 0.994 & \textbf{1.000} \\
TF$_{merge}$+B & - & - & - & - & - & - & - & - & - & 0.365 & 0.988 & \textbf{1.000} & 1.845 & 0.989 & 0.994 & - & - & - & - & - & - \\
TF$_{split}$+B & - & - & - & - & - & - & - & - & - & - & - & - & - & - & - & 0.660 & 0.989 & \textbf{1.000} & 17.220 & 0.994 & \textbf{1.000} \\
TF+B+P        & 2.605 & 0.979 & 0.996 & 0.720 & 0.990 & 0.996 & 0.695 & 0.986 & \textbf{1.000} & 0.195 & 0.988 & \textbf{1.000} & 1.485 & 0.988 & 0.994 & \textbf{0.000} & 0.996 & \textbf{1.000} & 0.840 & \textbf{0.996} & \textbf{1.000} \\
\bottomrule
\end{tabular}
}
\label{tab:ged-ss-stable}
\end{table*}
\begin{table}[ht]
\centering
\small
\setlength{\tabcolsep}{2pt}
\renewcommand{\arraystretch}{1.05}
\caption{Plan refinement execution accuracy on Stepwise Math Reasoning by operation type and feedback condition. Bold and underline mark the best and second-best.}
\begin{tabular}{l|c|c|c|c|c|c}
\toprule
\textbf{Operation}  & \makecell{\textbf{Add} \\ \textbf{Node}} & \makecell{\textbf{Change} \\ \textbf{Desc.}} & \makecell{\textbf{Change} \\ \textbf{Agent}} & \makecell{\textbf{Merge} \\ \textbf{Sequential}} & \makecell{\textbf{Merge} \\ \textbf{Parallel}} & \textbf{Avg.} \\
\hline
GF         & \textbf{0.860} & 0.860 & \textbf{0.860}   & \underline{0.860} & \textbf{0.760} & \textbf{0.840} \\
TF        & 0.128 & \underline{0.900} & 0.800   & 0.857 & 0.020 & 0.545 \\
TF$_{merge}$ & - & - & - & \underline{0.860} & 0.000 & 0.434 \\
TF+P        & \underline{0.378} & 0.880 & 0.720  & \textbf{0.900} & \underline{0.400} & \underline{0.657} \\
\hline

GF-to-DM        & 0.268 & \textbf{0.907} & \textbf{0.860}  & 0.370 & 0.167 & 0.553 \\
\hline
\multicolumn{7}{c}{\textbf{Flexible Boundary}}\\
\hline
TF+B        & 0.130 & 0.420 & 0.260   & 0.100 & 0.020 & 0.187 \\
TF$_{merge}$+B & - & - & - & 0.560 & 0.000 & 0.280 \\
TF+B+P        & 0.213 & 0.740 & 0.700  & 0.780 & 0.080 & 0.506 \\
\bottomrule
\end{tabular}
\label{tab:execution-accuracy}
\end{table}

\paragraph{Targeted feedback helps on plan stability but worsens in structural correctness}
\rev{Table~\ref{tab:ged-ss-stable} presents refinement performance across operation types for different feedback conditions.}
Targeted feedback improves plan stability compared to global feedback.
Because targeted refinement revises only the selected subgraph, it naturally yields higher plan stability: the non-target portion of the plan is largely preserved by construction.
Accordingly, all targeted conditions achieve higher stability than global feedback.
\textbf{GF-to-DM} shows a similar stability advantage. Since \textbf{GF-to-DM} converts a global feedback into a sequence of localized direct-manipulation operations, it also leaves untouched nodes unchanged and therefore achieves perfect stability, comparable to \textbf{TF}-based conditions.

However, the conditions differ in structural correctness. Targeted feedback does not consistently outperform global feedback: across most operation types, \textbf{TF} and \textbf{TF+P} produce higher GED than \textbf{GF}, indicating larger deviations from the gold plan topology.
\textbf{GF-to-DM}'s GED is between that of \textbf{GF} and targeted-feedback settings,
suggesting that edit-sequence revision can preserve plan stability while avoiding some of the structural drift introduced by free-form targeted rewriting.
For simpler operation types, such as changing the assigned agent, \textbf{GF-to-DM} can even achieve perfect GED, showing that constrained edit operations are especially effective when the intended repair maps cleanly onto a small set of direct manipulations.

\paragraph{Auto-merge preserves structure more reliably than auto-split}
Across operation types, the LLM-assisted \textbf{DM$_{high}^+$} variants largely mirror the targeted-replanning trends observed in \textbf{TF}, but with asymmetric effects on structural deviation. 
In particular, applying auto-merge based on \textbf{TF} tends to be structure-preserving: it reduces the GED introduced by targeted replanning, or yields a GED comparable to \textbf{TF} baseline. 
In contrast, auto-split is structurally more disruptive and typically increases GED relative to \textbf{TF}, indicating that split operations introduce larger topology changes than merge operations. 

\paragraph{Direct plan regeneration is most robust on complex structural revisions}
Table~\ref{tab:execution-accuracy} reports  execution accuracy across operation types on Stepwise Math Reasoning tasks. 
Overall, global replanning achieves the highest average accuracy (0.840) across all plan regenerations, substantially outperforming structural operation generation with an average accuracy of 0.553 (0.287 difference).
This suggests that the failure of structural operation refinement arises from incomplete recovery from $p_{\textit{initial}}$ to the $p_{\textit{gold}}$ structure, which then propagates to incorrect final answers during execution.

Here, robustness refers to how reliably a strategy produces a globally consistent plan and executable plan when the required revision involves structural updates, rather than a single parameter change.
Consistent with this definition, the performance gap is concentrated in structural edit operations. 
For operations that require coordinated topology changes (Add Node and Merge), \textbf{GF-to-DM} is markedly worse than \textbf{GF} (\ie, Add Node: 0.268 vs.\ 0.860; Merge Sequential: 0.370 vs.\ 0.860; Merge Parallel: 0.167 vs.\ 0.760).
These sharp drops indicate that step-wise edit sequences are less robust when a refinement requires multiple node/edge updates: small mistakes in early operation can compound, leading to a plan that executes incorrectly even if the plan structure appears plausible.
In contrast, for more localized semantic edits, such as Change Description and Change Agent, \textbf{GF-to-DM} is competitive with (and sometimes outperforms) \textbf{GF} (\ie, Change Desc.: 0.907 vs.\ 0.860 and Change Agent: 0.860 tie).

\paragraph{Edit-sequence revisions trade reliability for interpretability.}
In terms of interpretability, \textbf{GF-to-DM} provides an explicit audit trail of operations, enabling users to inspect what changed at each step and potentially diagnose where a revision went wrong.
By comparison, \textbf{GF} outputs a revised plan in a single pass without revealing intermediate decisions.
These results highlight a key trade-off: edit-sequence generation offers higher interpretability (auditable, step-wise operations) and can be effective for localized edits, but direct regeneration is more reliable for complex structural changes that require globally consistent revisions.
\label{sec:experiment}

\section{Discussion}
\subsection{Design Suggestions}
Our results reveal a pattern that richer interaction types shift user effort from authoring to verification and integration.
In the user study, advanced interactions (particularly LLM-assisted structural edits) reduced perceived effort and improved ease of use, yet they did not improve final plan quality.
Complementing this, our LLM plan-revision experiments show that global feedback yields the strongest overall performance.
However, effective global feedback assumes that users can reason about the full plan state and articulate precise guidance: an interaction style that participants were less inclined to use or prefer. 
Instead, participants often relied on hybrid workflows, combining targeted feedback for coarse replanning with direct manipulation for local refinements.
This strategy partially mitigates the performance gap of targeted feedback by offloading global-coherence corrections to human edits.
These findings suggest that the primary bottleneck in collaborative plan revision is not merely the efficiency of plan authoring, but the verification and integration of edits: users must ensure that modifications preserve global dependencies, maintain boundary compatibility, and guarantee execution validity.
These findings point to the following design suggestions:

\paragraph{Verification and integration support}
Our findings suggest that the main barrier to higher plan quality is not generating edits faster, but making them easy to trust and integrate.
To reduce ``blow-up'' concerns and verification burden, systems should shift support toward verification and boundary management:
Preview how a proposed visualized edit changes nodes and edges at reintegration points, and validate compatibility before applying revisions.
In addition, lightweight, risk-triggered checks (such as orphan node detection and input/output mismatch warnings) can counter verification decay and help users confirm the plan without heavy manual inspection.

\paragraph{Proactive context-aware interaction guidance}
Since users naturally combine targeted feedback with direct manipulation, interactive planners should make proactive context-aware guidance a first-class feature across both modes.
When users select nodes, the system can recommend targeted feedback prompts tailored to the selected region and its boundary contracts (\ie, ``split this node into smaller sub-steps'', ``duplicate these nodes and parallelize them with the existing branch'').
After replanning, the system should immediately surface likely follow-up fixes, \eg, missing variable bindings, orphan nodes, boundary mismatch, and offer one-click repairs. 
Finally, the interface can suggest the next best action to reduce iteration overhead and prevent verification fatigue.

\paragraph{Domain-specific output inspection}
While the planning and editing process can remain domain-agnostic, systems should provide domain-specific views for inspecting intermediate outputs. 
Our current interface assumes that node outputs can be summarized as short text, but richer workflows may produce code, tables, images, long-form text, or web search results that are difficult to verify within a node card. 
Future systems should therefore pair a general DAG-based planning shell with dedicated output panels, such as code diffs, rendered image previews, table viewers, or side-by-side long-text comparisons, so users can verify both structural changes and domain-specific execution results.

\subsection{Limitations}
\label{sec:limitations}

Despite providing insights into human–LLM collaborative planning, our work has several limitations. 
First, our co-planning paradigm was restricted to a single user operating on an orchestrated MAS with a fixed set of specialized agents, leaving unexplored other interaction paradigms and MAS configurations, such as multi-user collaboration or cooperative and more decentralized agent settings. 
Second, our user study involved a small sample (n=13) drawn from a single research lab, with most participants already possessing strong LLM expertise, which may limit generalizability to broader and more diverse populations. 
Third, participants’ verification effort declined over time due to fatigue, potentially biasing plan quality measures, and our evaluation metrics may not fully capture more nuanced aspects of robustness or interpretability.
Fourth, the artificially curated tasks, while multi-agent and multi-step, were relatively constrained in complexity and may not reflect highly dynamic or open-ended planning scenarios that could most benefit from human steering (\eg, domain-specific or personal workflows, or long-horizon tasks such as deep research). 
\rev{Finally, the initial erroneous plans used for our main experiments were synthetically constructed. While our supplementary experiment with naturally-occurring planner failures (Appendix~\ref{app:natural-faulty-plans}) suggests our key findings generalize, larger-scale validation across diverse planner models and failure distributions remains future work.}
\label{sec:dis}

\section{Conclusion}
\rev{In this paper, we formalized a design space for human-LLM co-planning along three axes (mode, scope, level) and implemented it in \tool. Through a user study and controlled benchmark, we found that users dynamically construct hybrid workflows, alternating across interaction types to navigate an effort-control-risk trade-off; while LLM revision is most robust under global feedback, targeted feedback is essential for preserving structural stability. 
Together, these contributions move human-LLM co-planning toward more transparent, controllable, and trustworthy multi-agent systems.}
\label{sec:conclu}


\bibliographystyle{ACM-Reference-Format}
\bibliography{bib/ref}

\clearpage
\appendix

\section{Implementation Details}
\label{appendix:imple}

Our prototype \tool is implemented as a web application with a React frontend communicating via FastAPI, and a Python backend handling core logic. The system is driven by an LLM-based planner and execution agents, each specialized for different tasks.
In our experiments, the planner and execution agents are configured to use GPT-4o and GPT-4o-mini, respectively.

\subsection{Planner Prompts}
\label{appendix:planner_prompts}
We provide the prompts used by \tool's planner for plan generation, replanning based on text feedback, subgraph replanning based on targeted text feedback, auto-split, and auto-merge.

\begin{imageonly} \begin{promptbox}[app:prompt_planning]{Prompt: Plan Generation}
\begin{lstlisting}[style=promptcontent]
You are an expert at breaking down tasks for planning. 
You will only have access to these agents:

[code] — For PURE coding tasks:
  - Implementing or modifying code to meet a spec (e.g., parse/transform text/JSON/CSV, write functions, simulate small programs, validate formats).
  - Algorithmic procedures best expressed as code (loops, data structures, regexes, parsing).
  - Debugging code or reorganizing/refactoring code.
  - NOT for mathematical derivations or symbolic reasoning. If a node mixes coding + math, split them: use [math] to derive, then [code] to implement.

[math] — For mathematical reasoning nodes:
  - Solving sub-problems in math: derive formulas, manipulate expressions, do case analysis, solve equations/inequalities, compute with given numbers.
  - Identify and restate conditions/variables; produce machine-evaluable expressions or numeric results where inputs are available.
  - Do NOT write or reason about code here. Keep it math-only.
  - The task MUST be a variable-template instruction (no concrete numbers). Use variable names only.
  - Never include numeric literals, percent symbols (%), or signs in math tasks; bind all given numbers in the node's input values.
  - Every variable listed in "variables" field in the input list MUST appear verbatim in the task description text.
  - The task description MUST NOT reference any other nodes.
  - For [math] nodes:
    • For each v in variables field in the input list, the task MUST contain v as a standalone token (exact match).
    • Reject tasks where a near-variant appears (e.g., "total sale") instead of the exact variable name (e.g., total_sales).
    • If a quantity is needed but not bound, create an upstream node to bind it to a properly named variable, then reference that exact name.

[search] — For retrieving specific factual knowledge from the Web (history, sports, culture, geography, medicine, science, etc.).

[commonsense] — For everyday reasoning that does not require Web retrieval (e.g., comparing magnitudes, widely-known facts, straightforward logical checks).

====================
GLOBAL INSTRUCTIONS
====================
Given a complex question or task, generate a structured, step-by-step plan to solve it.

Each node MUST follow this JSON schema:
{
  "id": <int>,
  "task": "<a complete, self-contained instruction using ONLY variable names from this node's inputs; never include discovered values.>", // Do not mention any other nodes in the task description!
  "agent_name": "<agent_name>",    // choose exactly one agent; if more than one seems needed, split into multiple nodes
  "input": [{"variable": "<variable_name>", "value": "<value>"}],  // bind given constants here; leave '' if unknown. 
  "output": ["<output_key>"],
  "prereq": [<node_id_1>, <node_id_2>, ...]
}

Also output the dependency edges (a plan graph). Each edge indicates that an output from one node is used as an input name in another node:
{
  "src_node": <source node id>,
  "dest_node": <destination node id>,
  "src_output": "<output key from source>",
  "dest_input": "<input key expected by destination>"
}

=============
PLANNING RULES
=============
1) Break the problem into independent, atomic nodes.
2) Each node is an INSTRUCTION only—describe what must be done, not the result.
   - You may include constants ONLY if they appear explicitly in the original problem statement.
   - Do not invent, look up, or leak unknown values into the plan; such values must be produced by earlier nodes or via [search].
   - Do NOT mention any other nodes in the task description. 
   - Do NOT mention any other nodes in the task description. 
   - Do NOT mention any other nodes in the task description. 
3) A single agent must be able to complete each node using ONLY:
   - the node's instruction,
   - the specified agent, and
   - outputs from its prereqs.
4) Do NOT reference “the original question” inside nodes. Rewrite what's needed directly into each node's instruction.
5) Use exactly one agent per node in the "agent_name" field. If multiple agents seem required, split the node.
6) Include any necessary variable names directly in the instruction so the executing agent has everything it needs. Use snake_case for output variable names.
7) Produce a valid DAG:
   - No isolated nodes.
   - A single sink node (the node with the highest id) is the final output node.
8) Edges:
   - Only create edges for actual data dependencies (where a later node's input name matches a prior node's output variable name).
   - Every edge must point from an existing output to a named input expected by the destination node.
\end{lstlisting}

\promptdivider
\begin{lstlisting}[style=promptcontent]
<given task>
\end{lstlisting}
\end{promptbox} \end{imageonly}

\begin{imageonly} \begin{promptbox}[app:prompt_replan]{Prompt: Replanning}
\begin{lstlisting}[style=promptcontent]
<same system prompt as plan generation>
\end{lstlisting}

\promptdivider
\begin{lstlisting}[style=promptcontent]
A plan and user feedback are given to you. Your job is to fix the plan according to the user feedback.

Conversation History:
<conversation history>

Plan:
<entire plan>

User Feedback:
<user feedback>
\end{lstlisting}
\end{promptbox} \end{imageonly}

\begin{imageonly} \begin{promptbox}[app:prompt_subgraph_replan]{Prompt: Subgraph Replanning}
\begin{lstlisting}[style=promptcontent]
You are an expert at re-planning sub-graphs in task planning DAGs.  
You will be given:  
1. A selected sub-graph (a set of nodes and connecting edges) as the focus for replanning.  

Your goal is to regenerate ONLY the selected sub-graph nodes, while keeping the interface (inputs/outputs defined by edges connecting to outside nodes) fully consistent.  

====================
GLOBAL INSTRUCTIONS
====================
- Every new node generated inside the replanned sub-graph must use an id that is a negative integer.  
  (Examples: -1, -2, -3, ...).  
- Do NOT use the original numeric IDs for new nodes. Keep original IDs only for nodes outside the replanned sub-graph.  
- Maintain the same **input and output variables** on the boundary edges of the selected sub-graph so that upstream and downstream connections remain valid.  
- All **edges from/to nodes outside the sub-graph must remain unchanged** in terms of:  
  • Outside node IDs  
  • Variable names  
- Inside the replanned sub-graph you may:  
  • Add, remove, or restructure edges  
  • Split or merge tasks across nodes  
  • Introduce additional internal connections  
  as long as the boundary interface to outside nodes remains consistent.  
- Do not modify nodes or edges outside the selected sub-graph.  

Each replanned node must follow this JSON schema:
{
  "id": -1,   // Use negative integers (-1, -2, -3, ...) for all new nodes inside the replanned sub-graph
  "task": "<a complete, self-contained instruction using ONLY this node’s input variables. Do not mention other nodes.>",
  "agent_name": "<agent_name>",  // [code], [math], [search], or [commonsense]
  "input": [{"variable": "<variable_name>", "value": "<value>"}], 
  "output": ["<output_key>"],
  "prereq": [<id_of_other_node>, ...]  // Can be a negative ID (inside sub-graph) or an original node id (outside sub-graph)
}

Also output the dependency edges among the replanned sub-graph nodes:
{
  "src_node": <node id>,   // negative ID (-) if inside sub-graph, positive original ID if outside
  "dest_node": <node id>,  // negative ID (-) if inside sub-graph, positive original ID if outside
  "src_output": "<output key from source>",
  "dest_input": "<input key expected by destination>"
}

=============
PLANNING RULES
=============
1. **Boundary consistency:**  
   - Any variable appearing on incoming edges from outside the sub-graph must appear as an input in at least one replanned node.  
   - Any variable appearing on outgoing edges to outside the sub-graph must be produced as an output by at least one replanned node.  
   - Outside node IDs and boundary edge structures must remain exactly the same.  

2. **Atomic instructions:**  
   - Each node must remain atomic, executable by exactly one agent.  
   - Split tasks if multiple agent types would be required.  

3. **Self-contained tasks:**  
   - Node instructions must not reference other nodes or “the original question.”  
   - Use variable names verbatim from inputs/outputs.  

4. **Valid DAG:**  
   - No isolated nodes.  
   - Exactly one sink node inside the replanned sub-graph.  

========================
RESPONSE FORMAT (JSON)
========================
{
  "nodes": [ <list of replanned node objects> ],
  "edges": [ <list of replanned edge objects> ]
}
\end{lstlisting}

\promptdivider
\begin{lstlisting}[style=promptcontent]
A sub-graph plan and user feedback are given to you. You job is to revise the subplan based on user's feedback

Sub-graph Plan:
<selected sub-graph>

User Feedback:
<targeted user feedback>

Note: Must have the inputs/outputs interface defined by edges to connect to outside nodes.
\end{lstlisting}
\end{promptbox} \end{imageonly}

\begin{imageonly} \begin{promptbox}[app:prompt_auto_split]{Prompt: Auto-Split}
\begin{lstlisting}[style=promptcontent]
<same system prompt as subgraph replanning>
\end{lstlisting}

\promptdivider
\begin{lstlisting}[style=promptcontent]
A sub-graph plan is given to you. You job is to split the sub-graph into a new plan. 
Keep the interface (inputs/outputs defined by edges connecting to outside nodes) fully consistent.

Sub-graph Plan:
<selected sub-graph>

Note: Must have the inputs/outputs interface defined by edges to connect to outside nodes.
\end{lstlisting}
\end{promptbox} \end{imageonly}

\begin{imageonly} \begin{promptbox}[app:prompt_auto_merge]{Prompt: Auto-Merge}
\begin{lstlisting}[style=promptcontent]
<same system prompt as subgraph replanning>
\end{lstlisting}

\promptdivider

\begin{lstlisting}[style=promptcontent]
A sub-graph plan is given to you. You job is to merge the sub-graph into EXACTLY ONE node. 
Keep the interface (inputs/outputs defined by edges connecting to outside nodes) fully consistent.

Sub-graph Plan:
<selected sub-graph>

Note: Must have the inputs/outputs interface defined by edges to connect to outside nodes.
\end{lstlisting}
\end{promptbox} \end{imageonly}

\subsection{Execution Agents}
\label{appendix:executor_prompts}
We implement four agents capable of solving selected datasets, similar to \citet{kim2024husky}. Each execution agent follows a task-specific pipeline that combines LLM reasoning with structured tool invocation: a Python execution tool for the Code Agent, a SymPy-based calculator for the Math Agent, and Google Custom Search or Brave Search APIs for the Search Agent.
Note that the selection and design of these execution agents primarily serve as a proof of concept, demonstrating that any agent capable of producing structured outputs can be integrated into our system with minimal modification. We provide the prompts used for each agent below.

~

\subsubsection{Code Agent}
~
\begin{imageonly} \begin{promptbox}[app:prompt_code]{Code Agent Prompt: Code Generation}
\begin{lstlisting}[style=promptcontent]
Given the input question, the solution history that consists of steps for solving the input question and their corresponding outputs, and the current step that must be addressed to solve the input question, write code that solves the current step.
- Write the code in Python.
- Do not attempt to write code that directly answers the question. Write code that answers the given step.
- For math questions, utilize the 'pi' symbol and 'Rational' from the sympy package for $\pi$ and fractions, and simplify all fractions and square roots without converting them to decimal values.
- Example imports are provided below. Import any of these packages, as well as additional packages as needed.
- Only generate the code, do not include any other text.
- Print the result of the code
- Convert the value of the result in string format before printing to json format
- The result should in json format with keys as %s
import math
import numpy as np
import sympy
from datetime import datetime
from math import comb, gcd, lcm
from scipy.optimize import minimize
from sympy import symbols, Eq, solve, expand, factor, Matrix
from sympy.solvers.inequalities import solve_univariate_inequality
from sympy.core.relational import LessThan
---
Question: <given subtask>
Code:
\end{lstlisting}
\end{promptbox} \end{imageonly}

\subsubsection{Math Agent}
~
\begin{imageonly} \begin{promptbox}[app:prompt_math]{Math Agent Prompt: Expression Generation}
\begin{lstlisting}[style=promptcontent]
You are a calculator assistant. 
Your job is to convert a math sub-task into a calculator-ready arithmetic expression **using only numbers** and basic operators: +, -, *, /, **, and parentheses.
**DO NOT include any unknown variables** (e.g., "a", "b") in the output expression. Use only the provided input variable values if they are numeric.
If an expression **cannot be fully evaluated** with the given numeric inputs, return null for that variable.

### Input
Task: <given subtask>
Output Variables: <expected output variables>

## Reasoning Requirement
You must provide your reasoning in the "thought" field, explaining:
- How you interpreted the mathematical problem
- Which numeric values you used and why
- How you constructed each expression
- Why you couldn't form an expression (if applicable)

### Output Format
If an expression can be formed:
{"thought": "...", "output_results": [{"key": "...", "value": "..."}, ...]}
else:
{"thought": "...", "output_results": [{"key": "...", "value": null}, ...]}

**Again DO NOT include any unknown variables** (e.g., "a", "b") in the output expression. 
\end{lstlisting}
\end{promptbox} \end{imageonly}

\subsubsection{Search Agent}
~
\begin{imageonly} \begin{promptbox}[app:prompt_search]{Search Agent Prompt: Search Query Generation}
\begin{lstlisting}[style=promptcontent]
Given the input question, write a concise, informative Google Search query for obtaining information regarding the input question. Do not use quotation
---
Question: <given subtask>
Search query Output Format:
{"thought": "...", "output_format": [{"key": "...", "value": "..."}]} 
\end{lstlisting}
\end{promptbox} \end{imageonly}

\begin{imageonly} \begin{promptbox}[app:prompt_search]{Search Agent Prompt: Response Generation}
\begin{lstlisting}[style=promptcontent]
You are a rewrite agent. Given the search question, the search results from the Google search api, answer the search question with the information in Search Results. 
Do not use your own knowledge to answer the question. 
Remove redundant information that is irrelevant to the question.
Fill those information into a json format with keys as %s. If there is no information, fill in empty string.
---
Question: <given subtask>
Search results: <web search results>
Output Format:
{"thought": "...", "output_format": [{"key": "...", "value": "..."}]} 
Answer:
\end{lstlisting}
\end{promptbox} \end{imageonly}

\subsubsection{Commonsense Agent}
~
\begin{imageonly} \begin{promptbox}[app:prompt_commonsense]{Commonsense Agent Prompt}
\begin{lstlisting}[style=promptcontent]
You are a commonsense agent. You can answer the given question
with logical reasoning, basic math and commonsense knowledge.
Fill those information into a json format with keys as %s. If there is no information, fill in empty string.
---
Question: %s
Output Format:
{"thought": "...", "output_format": [{"key": "...", "value": "..."}]} 
Output: 
\end{lstlisting}
\end{promptbox} \end{imageonly}
\section{User Study Details}

\subsection{Task Questions}
We provide the 8 questions used in user study tasks in Table~\ref{tab:user-study-question}.

\aptLtoX{\begin{table*}[ptb]
\centering
\caption{Task questions used in the user study (\S~\ref{sec:user-study-task}). Each pattern has two paired questions designed for matched difficulty.}
\label{tab:user-study-question}
\renewcommand{\arraystretch}{1.8}
\begin{tabular}{
  >{\centering\bfseries\arraybackslash}m{3.2cm}
  m{13.5cm}
}
\toprule
{Type} &
{\centering \textbf{Question}} \\

\midrule

\multirow{2}{*}{Stepwise Math Reasoning}
 & Tom buys a Michael Jordan autographed trading card for \$30,000 and spends \$10,000 on PSA grading. After grading, the card's value increases by 210\% of the original purchase price. How much profit did he make? \\
 \cmidrule{2-2}
 & Josh decides to try flipping a house. He buys a house for \$80,000 and then puts in \$50,000 in repairs. This increased the value of the house by 150\%. How much profit did he make? \\
\midrule
\multirow{2}{*}{Multi-Hop Computation}
   & What is the distance in kilometers between the birthplaces of the NBA Sixth Man of the Year in 1995 and 2023? \\
   \cmidrule{2-2}
 & What is the distance in kilometers between the city of birth of the NBA Rookie of the Year in 2005 and 1987? \\
\midrule

 \multirow{2}{*}{Listed Retrieval \& Aggregation}
 & What is the average number of chapters in the \textit{Game of Thrones} series? \\
 \cmidrule{2-2}
 & What is the average number of chapters in the \textit{Harry Potter} series? \\
\midrule

\multirow{2}{*}{Top-K Retrieval \&\\Aggregation}
 & What is the top 5 university tuition mean on the East Coast for resident students? \\
 \cmidrule{2-2}
 & What is the top 5 university tuition mean on the West Coast for resident students? \\
\bottomrule
\end{tabular}
\end{table*}}{\begin{table*}[ptb]
\centering
\caption{Task questions used in the user study (\S~\ref{sec:user-study-task}). Each pattern has two paired questions designed for matched difficulty.}
\label{tab:user-study-question}
\renewcommand{\arraystretch}{1.8}
\begin{tabular}{
  >{\centering\bfseries\arraybackslash}m{3.2cm}
  m{13.5cm}
}
\toprule
{Type} &
{\centering \textbf{Question}} \\

\midrule

\multirow{2}{*}[-0.5\baselineskip]{\makecell{Stepwise Math\\Reasoning}}
 & Tom buys a Michael Jordan autographed trading card for \$30,000 and spends \$10,000 on PSA grading. After grading, the card's value increases by 210\% of the original purchase price. How much profit did he make? \\
 \cmidrule{2-2}
 & Josh decides to try flipping a house. He buys a house for \$80,000 and then puts in \$50,000 in repairs. This increased the value of the house by 150\%. How much profit did he make? \\
\midrule
\multirow{2}{*}[-0.5\baselineskip]{\makecell{Multi-Hop\\Computation}}
   & What is the distance in kilometers between the birthplaces of the NBA Sixth Man of the Year in 1995 and 2023? \\
   \cmidrule{2-2}
 & What is the distance in kilometers between the city of birth of the NBA Rookie of the Year in 2005 and 1987? \\
\midrule

 \multirow{2}{*}[-0.5\baselineskip]{\makecell{Listed Retrieval \&\\Aggregation}}
 & What is the average number of chapters in the \textit{Game of Thrones} series? \\
 \cmidrule{2-2}
 & What is the average number of chapters in the \textit{Harry Potter} series? \\
\midrule

\multirow{2}{*}[-0.5\baselineskip]{\makecell{Top-K Retrieval \&\\Aggregation}}
 & What is the top 5 university tuition mean on the East Coast for resident students? \\
 \cmidrule{2-2}
 & What is the top 5 university tuition mean on the West Coast for resident students? \\
\bottomrule
\end{tabular}
\end{table*}}

\subsection{Metrics}
\label{appendix:user-study-metrics}

\paragraph{Effectiveness.} Effectiveness captures whether the system supports producing correct and well-formed plans after collaborative revisions.
We use the following measures:\footnote{The effectiveness grading was conducted manually. One author first graded all cases using the task-specific rubrics, and a second author independently verified the annotations. Any disagreements were resolved through discussion until a consensus was reached.}

\begin{itemize}
    \item \textbf{Task Success (final outcome accuracy).} For each task, we assess whether the participant produced a correct final answer according to task-specific criteria. Because tasks differ in output type (exact numeric/string answer vs. computed estimates), we define per-task grading rules and apply them consistently across both conditions. 
    The full task-by-task criteria are detailed in Table~\ref{tab:task-success}.
    \item \textbf{Plan Completeness (plan quality score).} To evaluate the quality of the produced plan independent of the final answer, we score each plan using a manual rubric based on whether it contains the essential reasoning/retrieval steps required by the task. Each task is scored on a 0-3 scale, awarding one point per essential step present (up to three points). 
    The task-specific essential steps and grading rubric are detailed in Table~\ref{tab:plan-score}.
\end{itemize}

\paragraph{Efficiency.} Efficiency captures the time and interaction effort required to complete a task. We report: (1) \textbf{Task Completion Time} (from task start to final submitted answer), (2) \textbf{Interaction Frequency}, measured as the total number of individual user operations (\eg, text feedback, targeted feedback, node/edge edits, merge/split, undo/redo), and (3) \textbf{Conversation Turns}, defined as the number of discrete interaction–execution cycles (\ie, each cycle is a sequence of user actions culminating in a single node execution or an ``Execute All’’ run). We compute these measures from the interaction logs and compare them across conditions.

\begin{table*}[ptb]
\centering
\small
\setlength{\tabcolsep}{4pt}
\renewcommand{\arraystretch}{1.15}
\caption{Task success criteria for the user-study questions in Table~\ref{tab:user-study-question}, used in manual grading.}
\label{tab:task-success}
\begin{tabular}{@{}
>{\raggedright\arraybackslash}p{0.22\linewidth}
>{\raggedright\arraybackslash}p{0.15\linewidth}
>{\raggedright\arraybackslash}p{\dimexpr0.63\linewidth-4\tabcolsep\relax}
@{}}
\toprule
\textbf{Type} & \textbf{Output} & \textbf{Success Criterion} \\
\midrule

Stepwise Math Reasoning & Numeric/string & Exact match to the ground-truth answer after normalization (\eg, commas/whitespace). \\
\midrule

Multi-Hop Computation & Distance (km) & Final distance value is within $\pm 10\%$ of the reference answer. \\
\midrule

Listed Retrieval \& Aggregation & Number of chapters & Final average chapter count is within $\pm 10\%$ of the reference answer. \\
\midrule

Top-K Retrieval \& Aggregation & Tuition amount & Final value passes a manual reasonableness check (expected scale: tens of thousands USD). If an explicit formula is provided, it must reflect an average (sum $\div$ 5). \\

\bottomrule
\end{tabular}
\end{table*}

\begin{table*}[ptb]
\centering
\small
\setlength{\tabcolsep}{4pt}
\renewcommand{\arraystretch}{1.15}
\caption{Plan completeness rubric for the user-study questions in Table~\ref{tab:user-study-question}, used in manual grading. One point per essential step (max 3).}
\label{tab:plan-score}
\begin{tabular}{@{}
>{\raggedright\arraybackslash}p{0.22\linewidth}
>{\raggedright\arraybackslash}p{\dimexpr0.78\linewidth-2\tabcolsep\relax}
@{}}
\toprule
\textbf{Type} & \textbf{Essential Steps (Up to 3 Points)} \\
\midrule

Stepwise Math Reasoning &
(1) Set up the correct equation/relations; (2) perform the required intermediate computations; (3) produce the final answer from the computed values. \\
\midrule

Multi-Hop Computation &
(1) Identify the target player(s); (2) retrieve birthplace(s); (3) obtain coordinates and compute distance. \\
\midrule

Listed Retrieval \& Aggregation &
(1) Determine the scope (which books count under the stated condition); (2) obtain chapter counts (per book or total); (3) compute the average. \\
\midrule

Top-K Retrieval \& Aggregation &
(1) Identify a set of top universities; (2) retrieve tuition for each school; (3) compute the average. \\

\bottomrule
\end{tabular}
\end{table*}

\subsection{User Study Questionnaire}
\label{appendix:user-study-survey}

\subsubsection{Background Survey}
\begin{enumerate}
    \item \textbf{Your User ID:} \_\_\_\_\_\_\_\_\_\_\_\_\_\_\_
    \item \textbf{How familiar are you with Large Language Models (LLMs)?} \\
    1 (Not at all familiar) $\cdots$ 7 (Extremely familiar)
    \item \textbf{How frequently do you use LLMs in your work or daily life?} \\
    \begin{itemize}
        \item Never
        \item Rarely (A few times a month)
        \item Occasionally (A few times a week)
        \item Frequently (Daily use)
        \item Constantly (Integrated into most of my work/life)
    \end{itemize}
    \item \textbf{Which of the following LLMs have you used?} (Select all that apply) \\
    ChatGPT, Gemini, Claude, Llama, Other: \_\_\_\_\_\_\_\_\_\_\_
    \item \textbf{In what ways have you interacted with LLMs?} (Select all that apply) \\
    General user (chat-based tools), Work-related tasks, LLM-powered product development, LLM research/development, Other: \_\_\_\_\_\_\_
    \item \textbf{How well do you understand LLM limitations (hallucination, bias, context limits)?} \\
    1 (Not at all) $\cdots$ 7 (Extremely well)
    \item \textbf{I trust LLM-generated information to be accurate.} \\
    1 (Strongly Disagree) $\cdots$ 7 (Strongly Agree)
    \item \textbf{I usually verify information provided by an LLM before using it.} \\
    1 (Strongly Disagree) $\cdots$ 7 (Strongly Agree)
\end{enumerate}

\subsubsection{Baseline System Questionnaire (1 = Strongly Disagree, 7 = Strongly Agree)}\hfill\\

\textbf{Text Feedback}
\begin{enumerate}
    \item It was easy to provide textual feedback on the plan.
    \item My text feedback led to plan modifications that matched my intent.
    \item Providing textual feedback required a lot of mental effort.
\end{enumerate}

\textbf{Direct Manipulation (Basic Operations)}
\begin{enumerate}
    \item It was easy to modify the plan using direct manipulation operations (add/edit/remove nodes/edges).
    \item My direct manipulation edits led to plan modifications that matched my intent.
    \item Providing feedback by direct manipulation required a lot of mental effort.
\end{enumerate}

\textbf{Overall Satisfaction}
\begin{enumerate}
    \item Overall, I was satisfied with the resulting plan after using this system.
    \item I felt in control of how my feedback changed the plan.
\end{enumerate}

\subsubsection{\tool Questionnaire (1 = Strongly Disagree, 7 = Strongly Agree)}\hfill\\

\textbf{Text Feedback}
\begin{enumerate}
    \item It was easy to provide general textual feedback on the entire plan.
    \item It was easy to provide targeted textual feedback by selecting specific steps.
    \item My textual feedback (general or targeted) led to plan modifications that matched my intent.
    \item Providing textual feedback (general or targeted) required a lot of mental effort.
\end{enumerate}

\textbf{Direct Manipulation (Basic + Advanced Operations)}
\begin{enumerate}
    \item It was easy to modify the plan using manual direct manipulation operations (add, edit, remove, merge, split, duplicate).
    \item It was easy to use LLM-assisted direct manipulation operations (auto-merge / auto-split).
    \item My direct manipulation edits (manual or assisted) led to plan modifications that matched my intent.
    \item Providing feedback by direct manipulation required a lot of mental effort.
    \item I trusted the LLM-assisted operations (auto-merge or auto-split) to make appropriate changes.
\end{enumerate}

\textbf{Overall Satisfaction}
\begin{enumerate}
    \item Overall, I was satisfied with the resulting plan after using this system.
    \item I felt in control of how my feedback changed the plan.
\end{enumerate}

\subsubsection{Post-Experiment Survey}
\begin{enumerate}
    \item \textbf{Preferred interaction mode for structural plan modifications:} \\
    Text Feedback, Targeted Text Feedback, Manual Direct Manipulation, LLM-Assisted Direct Manipulation
    \item \textbf{Preferred interaction mode for semantic or ambiguous modifications:} \\
    Text Feedback, Targeted Text Feedback, Manual Direct Manipulation, LLM-Assisted Direct Manipulation
    \item \textbf{Overall preferred interaction mode for plan modifications:} \\
    Text Feedback, Targeted Text Feedback, Manual Direct Manipulation, LLM-Assisted Direct Manipulation
    \item \textbf{Aspects of providing feedback that were most helpful or frustrating:} \_\_\_\_\_\_\_\_\_
    \item \textbf{When did you prefer using text vs. direct manipulation? Why?} \_\_\_\_\_\_\_\_\_
    \item \textbf{Would you consider using this system in daily or work planning tasks? Why or why not?} \_\_\_\_\_\_\_\_\_
    \item \textbf{Additional comments:} \_\_\_\_\_\_\_\_\_
\end{enumerate}

\subsection{Detailed Results}
\label{appendix:user-study-detail-results}

\subsubsection{RQ 1: How do advanced interaction types (\textbf{TF, DM$_{high}$}) affect user performances in terms of effectiveness and efficiency?} 
We evaluate each final plan completed by users using the task success and plan completeness rubrics described in \S~\ref{perform-metrics}.
In the main task, each participant completed 8 tasks.
Given our sample size (n=13) and within-participant design, we focus on paired comparisons on the shared tasks and report descriptive statistics and within-participant differences rather than relying on large-sample significance testing.

\paragraph{H1: Advanced interaction types (\textbf{TF, DM$_{high}$}) produce better quality plans than basic ones (\textbf{GF, DM$_{low}$}).}
To test H1, we compare plan quality between advanced types and basic types using rubric-based measures (task success and plan completeness).
On average, participants produced higher-quality plans using the basic types than using the advanced types (task success: 0.712 vs.\ 0.635; plan completeness score: 2.904 vs.\ 2.808).
These comparisons do not support H1, suggesting that advanced features did not help users improve the plan and produce better quality.

\paragraph{H2: high-level structural interactions (\textbf{DM$_{high}$}) are faster with lower cognitive load than low-level structural interactions (\textbf{DM$_{low}$}).} 
To test H2, we compare the baseline system (\textbf{DM$_{low}$}) and \tool (\textbf{DM$_{high}$}) on three self-contained outcomes:
(1) efficiency (task completion time), (2) mental effort (NASA-TLX), and (3) matched intention (7-point self-report) that captures how well the system output aligned with participants’ intended edits.

\textit{Efficiency.} Participants completed tasks slightly faster with \textbf{DM$_{low}$} than with \textbf{DM$_{high}$} (8.163 vs.\ 8.256 minutes on average). 
\textit{Matched intention.} Participants reported slightly higher matched intention for \textbf{DM$_{low}$} than for \textbf{DM$_{high}$} (6.385 vs.\ 6.231).
\textit{Mental effort.} Participants reported higher cognitive load with \textbf{DM$_{low}$} than with \textbf{DM$_{high}$} (4.385 vs.\ 3.923).
These descriptive comparisons partially support H2: \textbf{DM$_{high}$} reduced cognitive load, but it was slightly slower and matched user intent marginally less than \textbf{DM$_{low}$}.
\smallskip

To further understand participants' perceptions of semantic text feedback relative to structural interactions, we compare these two interaction modes within each system. 

In the baseline system, participants reported that \textbf{GF} matched their intention less than \textbf{DM$_{low}$} (5.615 vs.\ 6.385), while requiring less mental effort (4.000 vs.\ 4.385). Participants also reported greater ease-of-use for \textbf{GF} than for \textbf{DM$_{low}$} (5.692 vs.\ 5.308). 
These results suggest that in the baseline system, semantic text feedback felt easier and less effortful, but at the cost of lower perceived alignment with users' intent.

In \tool, participants reported a similar pattern: text feedback (\textbf{GF+TF}) matched their intention less than structural edits (\textbf{DM$_{low}$+DM$_{high}$}) (5.769 vs.\ 6.231), while requiring less mental effort (3.154 vs.\ 3.923). 
Participants also rated the ease of use of additional interaction options.
LLM-assisted \textbf{DM$_{high}^+$} received the highest ease-of-use rating (6.000), followed by \textbf{GF} and \textbf{TF} (both 5.846), while manual \textbf{DM} received the lowest ease-of-use rating (5.231).
These comparisons suggest that semantic interaction consistently reduced mental effort but was perceived as less aligned with users' intent than structural interaction; notably, LLM-assisted \textbf{DM$_{high}^+$} was rated as the easiest option in \tool.

\paragraph{H3: Targeted text yields fewer conversational turns than general chat.}
To test H3, we compare interaction length between the baseline system and \tool using two measures: (1) the number of conversational turns and (2) the total number of editing interactions (text or DM operations).
Participants had more conversation turns in the baseline system than in \tool (3.615 vs. 3.404), but they performed fewer interactions in the baseline system than in \tool (29.750 vs. 45.357).
These comparisons support H3: the target replanning reduced conversational turns compared to general chat, while its richer feature set led participants to perform more operations during editing.

\subsubsection{RQ 2: How do users choose among interaction types for different plan modifications, and what recurring patterns emerge in collaborative refinement?}

\paragraph{Users did not ``pick a type''; they assembled hybrid workflows per iteration.}\label{RQ2-1}
In the post-task survey, overall interaction type preferences were broadly distributed across two structural types and two semantic types (\textbf{DM$_{high}^+$}: 5; \textbf{TF}: 4; \textbf{GF}: 4; \textbf{DM}: 0).
Importantly, \textbf{DM} was not selected as a primary type, but it was still used. 
Participants used \textbf{DM} for modifications that required fine-grained control, especially in semantic nuance and ambiguous cases where they wanted precise local changes (\S~\ref{RQ2-2}).
In practice, participants rarely stayed within one type throughout iterations. 
Instead, they assembled hybrid workflows that alternated between making larger changes via semantic interactions and repairing local details via structural interactions. 
A common sequence was: (1) review and/or execute the current plan, (2) apply targeted or global text feedback to structurally revise the plan, (3) use manual or LLM-assisted DM to refine task descriptions or reconnect edges, and (4) re-execute to validate the updated plan.
These observations suggest that type selection primarily served as a sequencing strategy during collaborative refinement rather than as a stable personal preference.

\paragraph{Intentions for structural modification pulled strongly toward LLM-assisted DM; semantic modification were more mixed.}\label{RQ2-2}
We hypothesized that users intending structural modifications (\ie, graph-level changes like merging and branching) would prefer LLM-assisted  \textbf{DM$_{high}^+$}, whereas users intending semantic/ambiguous modifications (\ie, changing intent or clarifying steps) would prefer \textbf{TF}.
This hypothesis was strongly supported for structural modifications but only weakly supported for semantic/ambiguous modifications.
For graph-level changes, participants reported preferring LLM-assisted DM (\textbf{DM$_{high}^+$}: 7; \textbf{GF}:4; \textbf{TF}: 2; \textbf{DM}: 0). 
In think-aloud and logs, participants used LLM-assisted  \textbf{DM$_{high}^+$} to rapidly decompose or consolidate structure. For example, users used auto-split to turn a single dense node into a well-formed subgraph or auto-merge to remove redundant nodes.
Several participants described auto-split and auto-merge as the most helpful features.
In contrast, when participants worked on semantic or ambiguous modifications, preferences were evenly distributed (\textbf{TF}: 4; \textbf{GF}:3; \textbf{DM$_{high}^+$}: 3; \textbf{DM}: 3).
Rather than converging on a single type, participants chose a different type depending on the amount of semantic change required.
For instance, clarifying an underspecified step or resolving a variable mismatch often required localized restructuring.
In practice, many semantic changes required localized structural edits (\ie, splitting a complex step into a subgraph and reconnecting inputs/outputs).
As a result, participants frequently paired \textbf{TF} (within a targeted region) with \textbf{DM} (to confidently perform fine-grained adjustments).

\paragraph{Type choice was mediated by an effort-control-risk trade-off.}\label{RQ2-3}
Participants implicitly treated \textbf{DM} as high-control but high-effort, LLM-assisted  \textbf{DM$_{high}^+$} as a structural leverage with bounded risk, \textbf{TF} as high semantic leverage with bounded scope, and the \textbf{GF} as the lowest authoring effort but the highest verification burden.
Across participants, type choice reflected an implicit trade-off between effort, control, and rewrite risk.
P8 mentioned that it was a ``trade-off between DM (know what will happen but more effort) and text (easy)''.
\textbf{DM} was consistently treated as the most controllable option, but also the most labor-intensive, including heavily manual editing, variable selection, edge connecting, and layout organization. 
Consistent with this, participants used \textbf{DM} primarily for local, low-risk operations, such as repairing edge connections, tightening input/output variables, or making brief description edits, especially when they already knew what needed to change (as P1 and P10 noted, \textbf{DM} felt preferable for minor modifications). 
\textbf{GF} reduced mechanical effort but introduced uncertainty. P12 framed \textbf{GF} as ``rolling a dice,'' especially for global-plan replanning.
P12 worried that global replanning could ``blow up their plan'' and therefore avoided replanning entire graphs.
LLM-assisted  \textbf{DM$_{high}^+$} occupied a middle ground by providing structural leverage with minimal instruction. 
Participants used auto-split for routine decomposition and as idea-seeking when uncertain how to proceed (P7).
These dynamics align with participants' interface requests aimed at reducing DM overhead and verification burden:
P1 requested auto-save for input/output edits; P8 suggested to have ``visual difference' after revisions; and P7 asked for loop/iteration support to reduce repetitive manual text editing.

\paragraph{Collaborative refinement showed two recurring rhythms: review-first versus execute-first.}\label{RQ2-4}
Participants diverged into two stable refinement rhythms that determined whether they executed or edited.
Most participants (10/13) adopted a review-first rhythm: they inspected the plan graph node by node, cleaned unnecessary nodes, and attempted to make the plan ``correct enough'' before executing it. 
A smaller group (3/13) followed an execute-first rhythm: they executed early to surface failures and used results to guide replanning. 
Despite this difference, both groups converged on the same iterative structure: broad edits via text feedback (targeted or global), followed by local DM patches, followed by re-execution.
Task context shaped where effort was put. 
Decomposition-heavy tasks (\ie, tuition/books) often elicited text-based feedback, while multi-hop questions often pulled users toward DM for duplication, edge wiring, and intermediate verification. 

\paragraph{Users used text feedback as an executable specification (mostly prescriptive), switching to diagnostic feedback when debugging.}\label{RQ2-5}
We analyzed 178 text feedback messages across two systems: the baseline system supported only \textbf{GF}, whereas \tool supported both \textbf{GF} and \textbf{TF}.
In the baseline system, participants provided 97 \textbf{GF}. 
In \tool, participants used \textbf{TF} slightly more often than \textbf{GF} (42 vs.\ 39).

Across both systems, participants tend to use \textbf{GF} when they anticipate plan-wide changes, such as large decompositions, the insertion of missing intermediate nodes, and revisions to the output variables.
In the \textbf{GF}, they often referenced explicit node IDs (\ie, ``copy node 1\ldots'', ``modify node 5\ldots'').
By contrast, when participants viewed the issue as localized, such as an input mismatch, a single node too complex, or steps needing duplication, they more often used the \textbf{TF} (\ie, ``split this node into X nodes'').

Feedback style was overwhelmingly prescriptive/instructive \\(169/178), where users provide commands to rewrite the plan graph (\ie, ``split'', ``connect'', ``use output from node X'').
This pattern occurred in both the \textbf{GF} and \textbf{TF}, indicating that participants treated replanning as specifying an algorithm rather than as providing high-level guidance.
The minority diagnostic feedback (9/178) occurred primarily during debugging: users first identified a correctness breakdown (\eg, incorrect formulas, incorrect variable semantics, or outputs that did not match the task). 
In these moments, users wrote feedback more like bug reports that named the mismatch and requested correction, sometimes followed by a prescriptive restatement of the intended computation.

\paragraph{Trust increased in-task but verification declined later in sessions: a trust-fatigue paradox that interacts with type choice}\label{RQ2-6}
Participants reported an increase in trust in LLM-assisted  \textbf{DM$_{high}^+$} from pre-task background LLM trust to in-task experience (3.620 to 5.538 on a 7-point scale).
They also self-reported a strong verification habit (M = 5.310, SD = 1.18).
However, observations in the user study revealed a consistent pattern of verification decay over time: participants often checked intermediate results early on, but reduced verification in the second half of the session as they rushed toward task completion.
Across participants, we observed three recurring verification behaviors:
(1) manual recomputation (especially for numeric tasks),
(2) inspection of intermediate node outputs and execution traces to diagnose failures, and
(3) lighter checks later in the session due to fatigue (\ie, accepting plausible-looking outputs without re-checking all dependencies).
This shift created a trust-fatigue paradox: trust in LLM-assisted  \textbf{DM$_{high}^+$} increased with experience, while verification effort decreased.
Verification behaviors were also influenced by type choice.
When participants anticipated higher risk or uncertainty, they tended to limit changes by using \textbf{DM} and \textbf{TF} to localize errors. 
When they were time-constrained, fatigued, or perceived the changes as low-risk, they often accepted LLM-generated results without checking.
Overall, the increase in trust did not translate into consistently rigorous verification. Instead, verification appeared as a constraint resource that diminished with fatigue, shifting participants toward faster but potentially riskier interaction strategies.
\section{Experiment Details}

\begin{table*}[t]
\centering
\small
\setlength{\tabcolsep}{4pt}
\renewcommand{\arraystretch}{1.15}
\caption{Integration success rates by operation type and feedback condition. ``--'' indicates not applicable.}
\label{tab:success-rates}
\setlength{\tabcolsep}{4pt}
\begin{tabular}{l c ccc cc|c cc c}
\toprule
\textbf{Operation} & \textbf{GF} & \textbf{TF} & \textbf{TF\textsubscript{split}} & \textbf{TF\textsubscript{merge}}& \textbf{TF+P} & \textbf{GF-to-DM} & \textbf{TF+B} & \textbf{TF\textsubscript{split}+B} & \textbf{TF\textsubscript{merge}+B}  & \textbf{TF+B+P} \\
\midrule
Add Node    & 1.000 & 0.870 & --    & --     & 0.920  & 0.820 & 0.900 & --    & -- & 0.945  \\
Change Task Description    & 1.000 & 1.000 & --    & --    & 1.000  & 0.860 & 1.000 & --    & --  & 1.000  \\
Change Node Agent      & 1.000 & 1.000 & --    & --     & 1.000 & 1.000  & 1.000 & --    & -- & 1.000 \\
Merge Sequential        & 1.000 & 0.995 & --    & 1.000 & 1.000 & 0.540 & 1.000 & --    & 1.000 & 1.000 \\
Merge Parallel    & 1.000 & 0.995 & --    & 0.985 & 1.000 & 0.720 & 1.000 & --    & 1.000  & 1.000 \\
Split Sequential        & 1.000  & 1.000 & 1.000 & --  & 1.000  & -- & 1.000 & 1.000 & --   & 1.000     \\
Split Parallel    & 1.000 & 1.000 & 1.000 & --   & 1.000 & --  & 1.000 & 1.000 & --  & 1.000     \\
\bottomrule
\end{tabular}
\end{table*}

\subsection{Detailed Integration Results}
\label{app:additional-integration-analysis}
\rev{Complementing the main results in Table~\ref{tab:integration-results-no-boundary}, Table~\ref{tab:success-rates} reports per-operation integration success rates across all conditions, including boundary-flexible variants.}

\subsection{Boundary Analysis}\label{sec:additional-boundary-analysis}
Complementing the analysis in \S~\ref{sec:exp:results}, we further analyze \textbf{TF+B} and \textbf{TF+B+P}, which extend \textbf{TF} and \textbf{TF+P}, respectively, by allowing flexible boundary interfaces during reintegration. 
Overall, integration success rates increase from \textbf{TF} to \textbf{TF+B}, \textbf{TF+P}, and \textbf{TF+B+P} (Table~\ref{tab:success-rates}). 
This pattern suggests that both boundary flexibility and full-plan context improve reintegration success, with full-plan context providing a slightly larger benefit than boundary flexibility alone. 
Combining both mechanisms, as in \textbf{TF+B+P}, yields the highest success rate among targeted replanning conditions. 
A similar trend appears for LLM-assisted merge operations: \textbf{TF$_{merge}$+B} $\geq$ \textbf{TF$_{merge}$}.

However, boundary flexibility also affects structural deviation (Table~\ref{tab:ged-ss-stable}). 
Within targeted replanning, allowing boundary interfaces to be updated can reduce integration failures, but it may also lower plan stability. 
Compared with boundary-freezing variants, \textbf{TF} and \textbf{TF+P}, boundary-flexible variants, \textbf{TF+B} and \textbf{TF+B+P}, can introduce additional structural changes because edits to the selected subgraph may propagate through its interfaces with the unchanged portions of the plan.
\rev{The auto-merge vs. auto-split asymmetry observed in \S~\ref{sec:exp:results} is amplified under boundary-flexible reintegration. For parallel split edits, both \textbf{TF+B} and \textbf{TF$_{split}$+B} show a large increase in GED (17.180 and 17.220) compared with other refinement settings (Table~\ref{tab:ged-ss-stable}).}
This suggests that allowing boundary interfaces to change during reintegration, when combined with split operations, can trigger revisions beyond the intended subgraph and produce plan topologies that diverge sharply from the gold structure.

\subsection{Naturally Occurring Faulty Plans}\label{app:natural-faulty-plans}
To evaluate revision on naturally occurring faulty plans, we prompted three planner models of varying capability (GPT-3.5 Turbo, GPT-4o-mini, and GPT-5-mini) to generate plans for the benchmark questions (\S~\ref{sec:bench-dataset}). 
Each model was run once per instance, yielding 200 plans per model. 

\rev{Stronger planner models produced faulty plans much less frequently. Out of 200, GPT-3.5 Turbo generated 51 faulty plans and 16 error plans, GPT-4o-mini generated 28 faulty plans (no error plans), and GPT-5-mini generated only 5 faulty plans (no error plans). 
Here, faulty plans preserve the expected plan structure but contain incomplete or incorrect planning content, whereas error plans deviate from the required format and cannot be parsed. 
Manual inspection showed that failures concentrated in edge-linking errors and input/output variable mismatches rather than in overall logical structure, with many plans containing multiple entangled defects.}

\rev{To assess revision quality in this setting, we conducted a repair evaluation on the 51 faulty plans from GPT-3.5 (the noisiest planner: 26 from Stepwise Math Reasoning, 4 from Multi-Hop Computation, 15 from Listed Retrieval \& Aggregation, 6 from Top-K Retrieval \& Aggregation), simulating \textbf{GF} and \textbf{TF} using a strong LLM (GPT-5.4). 
We report execution accuracy of revised plans. \textbf{GF} outperformed \textbf{TF} on three of four subsets (Stepwise Math Reasoning: 0.615, Listed Retrieval \& Aggregation: 0.133, Top-K Retrieval \& Aggregation: 0.333). Both methods failed on Multi-Hop Computation, and \textbf{TF} recovered almost no plans elsewhere (Listed Retrieval \& Aggregation: 0.067; otherwise 0).}

\rev{
This gap reflects two factors. First, \textit{LLM-simulated targeted selection and feedback are systematically lower-quality than LLM-simulated global feedback}: \textbf{TF} must both identify the faulty subgraph and formulate a localized repair, compounding error opportunities. Next, \textit{natural failures are often globally distributed across the plan rather than localized}, making them better suited to holistic revision than local repair. The Multi-Hop case illustrates the upper bound: when faults span the plan structure, even \textbf{GF} cannot fully recover.
This strengthens our main synthetic-benchmark finding (§\ref{sec:experiment}): naturally distributed defects, like our synthetic break types, favor holistic revision over local repair.}

\end{document}